\documentclass[conference]{IEEEtran}

\usepackage{amsmath,mathtools}
\usepackage{subfigure}
\usepackage{color}
\usepackage{algpseudocode}
\usepackage{algorithm}
\usepackage{ulem}
\usepackage{tabularx}
\usepackage{array}

\ifCLASSINFOpdf
\usepackage[pdftex]{graphicx}
\DeclareGraphicsExtensions{.pdf,.jpeg,.png} \else
\usepackage[dvips]{graphicx}
\usepackage{comment}
\fi

\usepackage{hyperref}
\usepackage{fancyhdr}

\fancypagestyle{firststyle}
{
   \fancyhf{}
   
   \fancyhead[C]{\scriptsize{This is the post-print of the following article: C. Suraci, S. Pizzi, D. Garompolo, G. Araniti, A. Molinaro and A. Iera, ``Trusted and Secured D2D-Aided Communications in 5G Networks," in Elsevier Ad Hoc Networks, doi: 10.1016/j.adhoc.2020.102403.\\Article has been published in final form at: \url{https://www.sciencedirect.com/science/article/pii/S1570870520307320}. \\\copyright~2021 Elsevier B.V. All rights reserved.}}
}

\newcommand{\change}[1]{\textcolor{blue}{#1}}

\begin{document}

\title{Trusted and Secured D2D-Aided Communications in 5G Networks} 

\author{
\IEEEauthorblockN{
C. Suraci\IEEEauthorrefmark{1}, S. Pizzi\IEEEauthorrefmark{1}, D. Garompolo\IEEEauthorrefmark{1}, G. Araniti\IEEEauthorrefmark{1}, A. Molinaro\IEEEauthorrefmark{1}, and A. Iera\IEEEauthorrefmark{2}\\}

\IEEEauthorblockA{
\IEEEauthorrefmark{1}CNIT/DIIES Dept., University Mediterranea of Reggio Calabria, Italy \\
e-mail: [chiara.suraci,sara.pizzi,david.garompolo,araniti,antmolin]@unirc.it\\
\IEEEauthorrefmark{2}CNIT/DIMES Dept., University of Calabria, Italy \\
e-mail: antonio.iera@dimes.unical.it\\}}

\maketitle

\begin{abstract}
The design of the forthcoming fifth generation (5G) system shall meet the severe requirement of managing an always increasing amount of traffic generated by both humans and machines, while guaranteeing data security.
Among the enabling technologies that will turn 5G into a reality, Device-to-Device (D2D) and Multicasting will certainly play a key role because of their capability to largely improve network resources utilization and to address emerging use cases requiring the delivery of the same content to a large number of devices.
D2D communications can help to improve traditional point-to-multipoint transmissions by reducing the multicast coverage area and exploiting properly selected relay nodes as data forwarders towards users with  worse channel conditions.
However, security issues are even more challenging for D2D connections, as data exchange happens directly between nodes in proximity.
To enhance performance and security of delivered traffic in 5G-oriented networks, in this paper we design SeT-D2D (Secure and Trust D2D), according to which trustworthiness inferred from both direct interactions and social-awareness parameters is exploited to properly select relay nodes. 
Main contributions of our research consist in the introduction of a model for the assessment of network nodes' trustworthiness and the implementation of security mechanisms to protect the data transmitted in D2D communications and the privacy of the involved users.
The conducted simulation campaign testifies to the ability of the proposed solution to effectively select relay nodes, which leads to an improved network performance.
\end{abstract}

\begin{IEEEkeywords}
5G networks, Device-to-Device (D2D) communication, Multicasting, Trustworthiness, Security.
\end{IEEEkeywords}

\thispagestyle{firststyle}

\IEEEpeerreviewmaketitle

\section{Introduction}
\label{sec:1}

Autonomous driving, tactile Internet, virtual and augmented reality, broadband and media everywhere, seamless connection of pervasive embedded sensors are just examples of the most attractive yet challenging use cases that the future fifth-generation (5G) network will turn into reality.
Device-to-Device (D2D) and Multicasting communications are expected to play a key role in helping to satisfy the demanding requirements of the foreseen use cases.
The former, thanks to its capability to offload cellular data traffic, enhance spectrum efficiency, and extend cell coverage \cite{D2D}; the latter, thanks to its capability to answer the increasing demand for multicast/broadcast multimedia services (such as, mobile TV, IP radio broadcasting, and video streaming and downloading) \cite{cisco}. 

Multicast transmissions represent the top notch solution to deliver group-oriented services, since multiple users can be fed through a single point-to-multipoint (PtM) transmission by exploiting the broadcast nature of the radio channel. In order to handle multicast and broadcast services over cellular networks, the 3rd Generation Partnership Project (3GPP) has standardized the evolved Multimedia Broadcast Multicast Service (eMBMS) \cite{embms}. 

Notwithstanding, still significant work must be done in view of an effective support of multicast/broadcast traffic in 5G networks. The traditional approach to multicast traffic delivery in cellular networks is the so-called Conventional Multicast Scheme (CMS) \cite{cms}, which guarantees perfect fairness, since all users are served and receive the same treatment, although it suffers from poor performance levels because users with good channel conditions are  constrained to the lowest data rate that cell-edge users can sustain. This is why in several works in the literature, D2D communications are exploited to improve the performance of multicast service delivery \cite{D2D-SF,d2d_multicast_1,d2d_multicast_2,tvt}. An interesting approach to select D2D transmitters, based on the channel conditions of the links between the D2D peers, for example, is proposed in \cite{D2D-SF}. 

The 5G system leveraging the mentioned solutions is, however, exposed to severe security threatens. Besides the magnified risk of security threats due to the huge number of 5G connected devices, D2D communications can raise further issues due to the direct connections between devices in proximity \cite{survey_d2d_sec1}. To make 5G a trustworthy multi-service platform, resilience, communication security, identity management, privacy and security assurance must be provided \cite{ss}.

In \cite{future_internet}, we proposed an algorithm that aims to enhance performance and security levels of a multicast CMS transmission. 
D2D clusters are formed to forward data towards users with the worst channel conditions, which are excluded from a direct multicast transmission from the gNodeB (gNB). Relay nodes (RNs), that forward multicast data sent by the gNB to users unable to directly receive them, are selected on the basis of their reputation resulting from their past interactions. In \cite{tob}, we tackled the same problem by proposing a reputation assessment mechanism in which users' trustworthiness is evaluated also taking into account the ``social'' reputation of the devices within the network. 

In the wake of our previous research, in this paper we introduce a mechanism to effectively deliver trustworthy multicast/broadcast traffic in 5G-oriented networks, named SeT-D2D (Secure and Trust D2D). The work in \cite{tob} is enhanced by introducing an accurate method to determine the trustworthiness of relaying nodes, which have to forward data towards users with worst channel conditions. In particular, the novelty is that two contributions are considered in computing node's trustworthiness:  
(i) the one derived from \textit{direct interactions}, thus based on the actual behavior of the node following its selection as a relay node,
(ii) and the other based on \textit{social trustworthiness} parameters, thus able to give a first evaluation even in absence of previous interactions. The idea of using both contributions stems from the fact that, especially in its early stages, it is very likely that a requester node has not interacted directly with a service provider (i.e., the RN) in the past; therefore, the latter is unable to properly evaluate its maliciousness due to the unavailability of information on its trustworthiness.
This issue is called \textit{cold start problem} in the literature, and implies that several interactions have to take place before a node is able to understand if another node is malicious or benevolent. To face this problem, we have designed a trustworthiness model which properly considers both contributions. 
Compared to \cite{tob}, we propose a non-trivial extension of the trust model that keeps into account several key parameters for estimating node's trustworthiness.

Another relevant contribution of this work consists in the employment of security techniques aimed at protecting the data transmitted via D2D communications and the privacy of users involved in the transmission. Namely, data transmitted in D2D are encrypted by using a symmetric encryption algorithm for which the two peers generate the secret key by leveraging the Diffie-Hellman Key Exchange (DHKE) protocol. Furthermore, another innovative factor compared to other existing works on D2D communications is that user identity privacy is preserved through the use of Subscription Concealed Identifier (SUCI) derived from the Subscription Permanent Identifier (SUPI) according to the 3GPP TS 33.501 \cite{3gpp}.

The remainder of the paper is organized as follows. \change{In Section \ref{sec:bground} we provide useful background information, at the basis of our proposal.} Section \ref{sec:rw} and \ref{sec:proposal} provide, respectively,  a brief overview of related works in the area of reference and a detailed description of the proposed SeT-D2D protocol, while Section \ref{sec:trust_model} describes the designed trustworthiness model. Results from our performance analysis are shown in Section \ref{sec:5}. Conclusive remarks are given in the last section. 

For the reader convenience, we summarize in Table \ref{tab:acro} the meaning of all acronyms used in the paper.

\begin{table}[htbp]
\centering
\caption{List of acronyms}
\label{tab:acro}
\scriptsize
\begin{tabular}{p{1 cm} p{6.7cm}}
\hline
ARPF & Authentication credential Repository and Processing Function \\
AKA & Authentication and Key Agreement \\
C-LOR & Co-Location Object Relationship \\
CMS & Conventional Multicast Scheme \\
CQI & Channel Quality Indicator \\
C-WOR & Co-Working Object Relationship \\
D2D & Device-to-Device \\
DHKE & Diffie-Hellman Key Exchange \\
EAP & Extensible Authentication Protocol \\
eMBMS & evolved Multimedia Broadcast and Multicast Service \\
eMBB & enhanced Mobile Broadband \\
gNB & gNodeB \\
HMAC & Hashed Message Authentication \\
IMSI & International Mobile Subscriber Identity \\
IoT & Internet of Things \\
MCS & Modulation and Coding Scheme \\
MG & Multicast Group \\
mMTC & massive Machine Type Communications \\
NFV & Network Function Virtualization \\
NR & New Radio \\
OOR & Owner Object Relationship \\
POR & Parental Object Relationship \\
PtM & Point-to-Multipoint \\
P2P & Peer-to-Peer \\
QoE & Quality of Experience \\
RAN & Radio Access Network \\
RN & Relay Node \\
SCC & Security Control Class \\
SIoT & Social Internet of Things \\
SDN & Software Defined Networking \\
SIDF & Subscription Identifier De-concealing Function \\
SOR & Social Object Relationship \\
SR & Security Realms \\
SUCI & Subscription Concealed Identifier \\
SUPI &  Subscription Permanent Identifier \\
UE & User Equipment \\
UDM & Unified Data Management \\
URLLC & Ultra Reliable and Low Latency Communications \\
3GPP & Third Generation Partnership Project \\
5G & Fifth Generation \\
\hline
\end{tabular}
\end{table}

\section{\change{Background}}
\label{sec:bground}

\subsection{\change{The 5G ecosystem}}

\change{The future 5G cellular network deployment will involve a radical change compared to previous generations. A plethora of innovative technologies are expected to be utilized, thanks to which novel use-cases will be supported. Based on use cases' requirements, three different service categories have been defined: \textit{(i) enhanced mobile broadband (eMBB)}, which includes services demanding very high peak data rates and improved Quality of Experience (QoE) (e.g., ultra high definition TV); \textit{(ii) massive machine-type communications (mMTC)}, which require connectivity for a massive number of energy-constrained Internet of Things (IoT) devices; \textit{(iii) ultra-reliable and low latency communications (URLLC)} supporting the most stringent requirements in terms of reliability and latency (e.g., V2X applications) \cite{5G_slicing,types_traffic}. Thus, ``flexibility'' shall be the key design concept in upcoming 5G networks to meet the diverse requirements of each category. In this view, Software Defined Networking (SDN), Network Function Virtualization (NFV), and Device-to-Device (D2D) communications are among the more promising enabling technologies that the 5G system will exploit \cite{5G-tec}. In particular, the D2D paradigm enables two devices to exchange data directly without going through the network. Countless benefits can be brought by D2D technology to cellular communications, among which high data rate, low latency,  extension of the network capacity,  energy saving, and network offloading stand out.
While on the one hand the 5G enabling technologies will allow the network to offer a greater number of services at improved conditions, on the other hand several new challenges must be faced, especially regarding network architecture and security.}

\subsection{\change{Security issues}}

\change{The DHKE algorithm is a well-known method for securely exchanging cryptographic keys, particularly suited to the generation of a secret between two peers since it requires only the exchange of information that, even if intercepted, does not allow an eavesdropper to obtain the secret. A critical vulnerability of DHKE is the man-in-the-middle attack, in which a malicious user impersonate a legitimate peer.}

\begin{figure}[ht]
\centering
\includegraphics[scale=0.3]{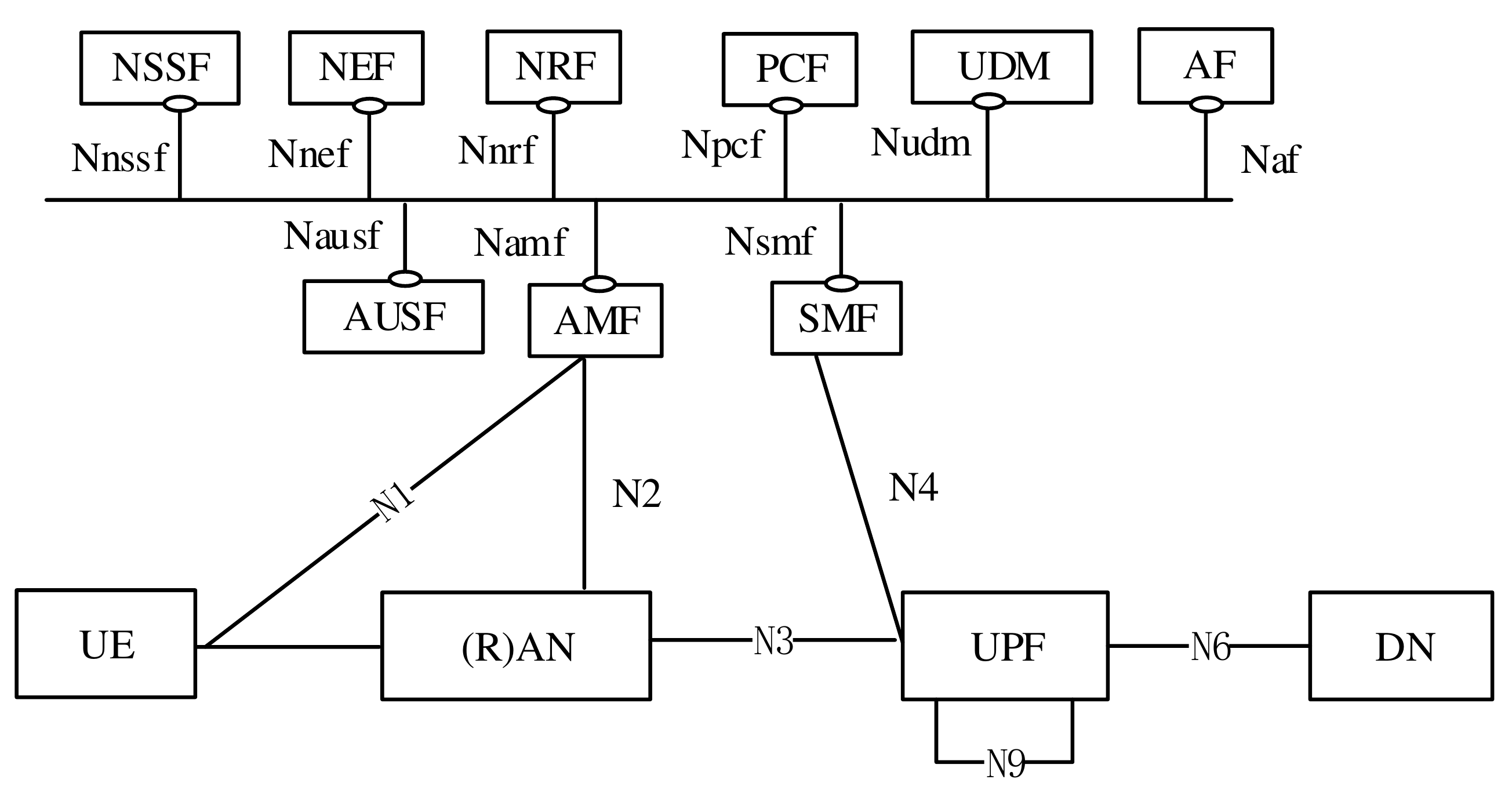}
\caption{The reference 5G architecture \cite{3gpp-arch}.} 
\label{fig:arch}
\end{figure}

\change{The concealment of the SUPI is another important security mechanism to ensure the protection of user privacy. The 3GPP specifies that, in order to preserve the subscribers privacy, \textit{the SUPI should not be transferred in clear text over NG-RAN (5G Radio Access Network) except routing information} \cite{3gpp}. \change{Thus, following 3GPP guidelines, SUCI has to be transmitted during authentication procedures \cite{3gpp}}. It is calculated by encrypting a part of the SUPI with the Home Network Public Key securely provisioned by the home network. Only user equipment (UE) and home network can de-conceal the SUCI to obtain SUPI; in the home network, the Subscription Identifier De-concealing Function (SIDF), located at the ARPF/UDM node (Authentication credential Repository and Processing Function/Unified Data Management), is responsible for this functionality. 
The reference architecture for this work is that described by 3GPP in \cite{3gpp-arch}, from which Fig. \ref{fig:arch} is taken, with the addition of security features and procedures described in \cite{3gpp}. As reported in \cite{3gpp}, only in three cases the protection of the SUPI may be missing: \textit{(i)} unauthenticated emergency sessions, \textit{(ii)} if the home network has set null-scheme to be used, \textit{(iii)} if the home network has not provisioned the public key needed to generate a SUCI.}

\subsection{\change{Evaluating nodes' trustworthiness}}

\change{In the research area addressing secure and trusted D2D communications, the social trustworthiness is considered a valuable means.}

\change{The literature uses the term \textit{trustworthiness} (in brief, trust) with a variety of meanings.
We have adopted one of its main interpretations: the \textit{reliability trust}. 
The reliability trust is the subjective probability by which an individual, \textit{A}, expects that another individual, \textit{B}, performs a given action on which its welfare depends.
Two fundamental concepts of the theory of trustworthiness are: (1) \textit{functional trust}, that is the trust in the ability of a node to provide services, and (2) \textit{referral trust}, that is the trust in the ability of a node to provide recommendations. In our scenario, referral trust is not considered, since the centralized nature of the network architecture makes it lose meaning. 
While trustworthiness is a general concept, in our specific scenario we are interested in the functional trust, which we refer to with the term \textit{service trust} for a more immediate understanding of the concept that it implies.
The service trust ($st$) is the trustworthiness parameter that we use to measure the trust in nodes' ability to provide services. 
Several trustworthiness models have been proposed for different technologies and architectures, ranging from Peer-to-Peer (P2P) to Internet of Things (IoT) and Social IoT (SIoT).}

\change{In the literature regarding the social trustworthiness, a fundamental parameter  is represented by the \textit{service integrity belief} (in brief, integrity), that allows to predict the behavior in future interactions and to obtain a trustworthiness value closer to Ground truth. In this way, the calculation of trustworthiness is not based only on the degree to which a node has satisfied past interactions (i.e., the competence), but also on the deviation in the degree of satisfaction of the recent interactions with respect to the remote ones (i.e., the integrity)}.

\change{The \textit{decay (or decaying) factor} indicates the expiry of data related to a given transaction. For the computation of the decay factor, some trustworthiness models consider only a number of transactions occurred after the one considered (current validity of the interaction or cardinal contribution), while others consider the time elapsed from the considered transaction (recency of the interaction or temporal contribution). However, the joint use of both contributions would allow a better estimate.}  

\change{In the proposed trustworthiness model, we also keep into account an \textit{indirect contribution} for assessing the trust of a node that, based on various parameters (i.e., relationship factor, centrality, and intelligence), allows to evaluate the trustworthiness of a node even in the absence of direct interactions. We will demonstrate that considering both a direct and an indirect contribution in trust computation allows to mitigate the cold start problem, as explained in Section \ref{sec:1}. More details about how the indirect contribution is measured are given in Section \ref{sec:trust_model}.}

\section{\change{Related Work}}
\label{sec:rw}

\subsection{\change{D2D for multicasting and related security issues}}

\change{Considered as one of the most important technologies for turning 5G into a reality, D2D communication has been the subject of a vast scientific literature \cite{D2D,D2D-SF,tvt}.
In \cite{D2D-SF}, D2D technology is presented as the means to improve the performance of a \textit{multicast} transmission via the establishment of direct communications between devices in proximity. Also in \cite{d2d_multicast_1}, D2D communications are considered highly valuable to improve content sharing in future 5G cellular networks, since they can offload traffic from cellular base stations and guarantee an efficient management of radio resources. The problem of power and channel allocation to multicast D2D communications is tackled in \cite{d2d_multicast_2} where two different solutions are proposed, both demonstrating the superiority of multicast D2D performance, especially in terms of throughput. 
In the mentioned research, not enough attention has been put on security, which is of paramount importance for future 5G networks. As evidence, a security architecture for 5G networks is defined in \cite{architecture_5G}, where network segments are logically divided into \textit{security realms (SR)}.
For each of them, security controls that could be implemented are classified in \textit{security control classes (SCC)}.}

\change{Privacy and security issues must be faced to definitely make D2D a successful technology for 5G systems. In \cite{security-privacy}, a thorough analysis on these problems is conducted. First, the difference between security and privacy concepts is defined. Then, the requirements to be satisfied in order to guarantee both security and privacy in D2D communications are listed, possible attacks identified, and solutions suggested. 
In \cite{survey-security}, security solutions proposed to improve D2D in 5G networks are analyzed. In addition, when dealing with D2D security requirements, threats, and solutions, this work hints at the role that social relationships could play in this context.}

\change{A secure data sharing strategy able to guarantee D2D privacy and security in LTE-A networks is proposed in \cite{seds}. The basic idea is to encrypt data transmitted over the D2D link by using a symmetric encryption algorithm and generating a private key, for data encryption and decryption, according to the DHKE protocol. The strength of this strategy lies in the intervention of the gNB as a trusted third party that can protect against malicious behaviors during the execution of the DHKE algorithm, thus overcoming the man-in-the-middle weakness. The SeT-D2D algorithm aims to overcome the gaps of this interesting research, providing a twofold contribution.
First, until a node is selected as a D2D transmitter, there is no information on its nature. Second, once a node is punished because of its malicious behavior, there is no way to redeem its reputation. The latter can be a problem when a node that usually performs  efficiently and trustfully is actually itself a victim of attack and temporary performs some malicious behaviors. From this time on, it is considered not eligible any more for the role of D2D transmitter and there is no way to rehabilitate it; if this node has the potential to be more efficient than others, an avoidable degradation of communications performance is observed.}

\subsection{\change{Social trustworthiness models}}

\change{The benefits of proximity-based mobile social networking are discussed in \cite{ometov}. In \cite{trust_indoor}, devices' characteristics (such as brand, type of friendship with other nodes, and computational capabilities) are considered to compose the social trustworthiness of each node, to the purpose of setting up trustworthy D2D communications between nodes with a good reputation. In \cite{reliable_relay}, a paradigm melting social trustworthiness and D2D communications is introduced to extend the coverage area of future 5G cells through the use of reliable relays. A framework that designs the caching based D2D communication scheme by taking social ties among users and common interests into consideration is proposed in \cite{wang}.}

\change{The Self-ORganizing Trust (SORT) model for P2P scenarios is proposed in \cite{Can}. Each peer computes the trustworthiness of other peers based on past interactions and recommendations, by using only locally available information. In particular, they measure the trustworthiness in providing services and in giving recommendations. To evaluate interactions and recommendations, they use importance, recentness, and peer satisfaction parameters. The introduction of the concept of integrity, as ``the level of confidence in predictability of future interactions'', is one of the major contributions of \cite{Can}. A flaw of the model in \cite{Can} concerns the decay factor, which only considers the cardinal contribution.}

\change{In \cite{Chen}, the authors present an access service recommendation scheme in a SIoT scenario which considers the social relationships among things. An energy-aware mechanism is also proposed to be utilized as a restrictive factor in trustworthiness evaluation. Even the recommendation is based not only on the past performance but also on the social relationship and on the energy status of nodes. Also in the model presented in \cite{Chen}, the decay factor is evaluated as a useful contribution to the computation of the trustworthiness of the network nodes. However, only the temporal contribution is considered, which leads to the previously discussed drawbacks.
Another well-known trust model in P2P systems is Peertrust \cite{Peertrust}. It presents a transaction-based feedback system built on three basic parameters and two adaptive factors: the feedback that a peer receives from other peers, the total number of transactions that a peer performs, the credibility of the feedback sources, the transaction context factor, and the community context factor. In \cite{Fuzzy}, the authors propose to use reputation as a means to establish the reliability trustworthiness of resources in a P2P scenario. In addition, an approach to manage and exchange reputations, based on the use of fuzzy techniques, is presented. A scalable, adaptive, and survivable trust management protocol in a dynamic IoT environment is designed and evaluated in \cite{Scalable}. Since entities in an IoT system are connected via the social networks of the entity owners, a social IoT based community of interest (CoI) is considered. The same authors, in \cite{Adaptive}, propose an adaptive trust management protocol for SIoT systems in which social relationships evolve dynamically among the owners of IoT devices. According to this approach, a social IoT application can adaptively choose the best trust parameter settings in response to changing IoT social conditions. Two other noteworthy works are \cite{Subj} and \cite{Obj}. 
In \cite{Subj}, each node computes the trustworthiness of its friends on the basis of its own experience and on the opinion of the friends in common with the potential service provider. To evaluate the trust level, a feedback system is used and the credibility, the centrality, and the intelligence of the nodes are combined. 
Differently, in \cite{Obj}, information about each node is distributed and stored by making use of a distributed hash table structure so that any node can exploit the same information.}

\subsection{\change{Main novelties introduced by the proposed trustworthiness model}}

\change{The designed trustworthiness model has the main aim to provide a \textit{fine-grain detection of service supply profile changes}. To achieve this goal, we leverage the following features:}
\begin{itemize}
    \item \change{Both \textit{direct} and \textit{indirect} contributions are considered, which are clearly separated and properly weighted. This allows greater control over the weight that can be associated (even dynamically) to each contribution. This choice derives from the opinion, broadly shared in the literature \cite{Can,Chen,Fuzzy,Scalable,Subj,Obj}, that this approach enables an easy runtime tuning of these two main contributions. Furthermore, the indirect contribution has been designed with the aim to mitigate the cold start problem.}
    \item \change{Each parameter contributing to the trust computation is \textit{properly weighted} to improve the adaptability to the considered scenario. This design principle has been used in \cite{Chen,Subj,Obj} for the indirect contribution, and in \cite{Can} for the direct contribution. As an example, weighting the integrity parameter allows to set possible punishments. Depending on the reference scenario, the central controller (e.g., the gNB) can decide the degree of harshness of the punishment given to a relay node providing a bad service. We demonstrate that, by correctly calibrating the integrity weight, we obtain the desired decrease of service trust. In this way, we reach a service trust value closer to the Ground truth.}
    \item \change{In the computation of the \textit{integrity} parameter, we propose to assess how short-term service opinion deviates from the long-term service opinion. A different approach is proposed in \cite{Can}, where the deviation from the average behaviour is calculated based on the deviation of the last satisfaction value (associated to the last transaction) from the average value of satisfaction computed over past transactions. The motivation behind our choice is that considering short-term service opinion and long-term service opinion allows us to obtain the expected service integrity belief and a service trust very close to the Ground truth.}
    \item \change{Both the cardinal and the temporal contributions are considered in the computation of the \textit{decaying factor} to its approximation to the Ground truth. In particular, we are able to better evaluate the temporal decay in scenarios where the considered interaction is occurred a long time before (long time interval), although it is among the last ones occurred (from a cardinal point of view). Furthermore, we are able also to better evaluate the cardinal decay in scenarios in which many interactions occurred after the one considered (short time interval), although this latter is recent. To the best of our knowledge, there are no models that jointly consider both contributions. In the literature, only the temporal contribution \cite{Chen} or only the cardinal contribution \cite{Can} is considered.}
\end{itemize}

\section{Providing Trusted D2D Communications in 5G-oriented Networks}
\label{sec:proposal}

In our reference scenario, devices located in the same coverage area are interested in downloading the same data. 
Main purpose of the proposed protocol is to optimize the data delivery from the provider to the network nodes, resorting to the establishment of \textit{secure and trusted D2D communications}. \textit{(1)} The \textit{D2D} implementation is finalized to support the multicast communication and to avoid that a mere CMS transmission penalizes all the receiving devices, by imposing to all disadvantageous transmission conditions caused by the cell-edge nodes. \textit{(2)} D2D communications are \textit{trusted} because an innovative model is introduced to assess the trustworthiness of potential D2D transmitters, in order to select trustworthy relay nodes. \textit{(3)} \textit{Security} in D2D is fixed, since the privacy of communicating devices is preserved and data transmitted are protected through the implementation of a symmetric encryption algorithm, for which the secret key is generated according to the DHKE protocol.

In detail, according to the proposed SeT-D2D protocol, devices with the best channel conditions are served directly by the gNB through a CMS transmission. Then, data are sent to cell-edge devices (i.e., those excluded from the multicast transmission because of their bad channel conditions) through secure D2D communications. Security is achieved by carefully selecting the D2D transmitters, referred to as relay nodes, and the implementation of some security mechanisms (such as DHKE,  Hashed Message Authentication Code (HMAC), and cryptography) aimed at guaranteeing confidentiality and integrity to data exchanged between devices in proximity. The selection parameters considered to choose the best RN for each D2D communication are: the channel quality indicator relative to D2D link (i.e., D2D CQI) and the trustworthiness of cell nodes. 

According to the 5G security architecture presented in \cite{architecture_5G}, we report in Table \ref{tab:sr_scc} the  SR-SCC mapping related to our use case (i.e., secure D2D data transmission). In the \textit{access network} domain,  the gNB shall be responsible for managing the credentials of users and protecting their privacy; moreover, identity of users has to be verified. In the \textit{application} realm, it is important to control data destinations since only registered users shall be affected; furthermore, the true identity of users shall not be revealed to the applications; therefore, fake identities could be used to reach them. For an effective \textit{management} of the overall network, some security mechanisms have to be implemented in order to provide information about the trustworthiness of the system. Finally, the \textit{UE} realm includes the ``other UE domains" that also cover D2D communications. Data transmitted during the direct communication between devices shall be confidential and their integrity  be protected. To this aim, security keys could be used to allow data encryption. In addition, the privacy of users shall be protected and the identity of the devices involved in the direct communications be checked to track any possible malicious behavior. As regards \textit{network} and \textit{infrastructure \& virtualisation}, i.e. the last two realms presented in \cite{architecture_5G}, the security mechanisms we propose do not affect these network segments, thus no security control class is reported for them.   

The proposed SeT-D2D protocol aims to implement the reported security controls for each security realm.
In the following, all steps of SeT-D2D are described in detail.

\begin{table*}[htbp]
\caption{Security realms-security control classes mapping for the reference use case.}
\centering
\begin{tabular}{p{5cm}p{5cm}p{6cm}}
\hline
\textbf{Security realms (SR)}	& \textbf{Security control classes (SCC)}   & \textbf{Security control examples}\\
\hline
\textsc{Access Network} 			& Authentication, Identity and Access Management, Privacy		&Credentials management, privacy protection, identities checks.   \\
\hline
\textsc{Application}			& Identity and Access Management, Privacy			& Data destinations controls, use of fake identities for privacy protection. \\
\hline
\textsc{Management} & Trust and Assurance &Knowledge of system trustworthiness.\\
\hline
\textsc{User Equipment} & Confidentiality, Integrity, Authentication, Privacy, Non-Repudiation & Security keys management, data encryption, D2D users identities checks, D2D users privacy protection. \\
\hline
\textsc{Network}			& 			& \\
\hline
\textsc{Infrastructure \& Virtualisation}			& 			& \\
\hline
\end{tabular}
\label{tab:sr_scc}
\end{table*}

\subsection{Multicast service delivery notification} 

Initially, the gNB announces the multicast service. It invites interested users to form the multicast group (MG) by registering to the network.

\subsection{Registration \& Authentication} 

Users must register to enjoy the multicast service notified by the gNB. Mutual authentication between UE and network is a primary requirement to ensure the security of both. Thus,  we assume that the authentication procedure described by 3GPP in \cite{3gpp} is implemented in our SeT-D2D algorithm. According to \cite{3gpp}, Extensible Authentication Protocol for Authentication and Key Agreement' (EAP-AKA') and 5G AKA are the protocols supported for the mutual authentication between UE and network. In particular, the ARPF/UDM node selects the proper authentication method for the user based on its SUPI and the subscription data. As already mentioned in the previous sections, the subscriber privacy is protected through the concealment of its permanent identifier to eavesdroppers on the air interface; the SUCI is transmitted in its place for this purpose. The likelihood of the occurrence of a SUPI catching attack is much lower than that of IMSI catching \cite{jover19}, nonetheless 3GPP states that, in some cases, SUPI could be exposed to attackers (see Section \ref{sec:rw}). For the scope of this work, we assume that users are always identified by their SUCI in the air interface, so the privacy of their identity is protected.

\subsection{CQI collection}

The network needs information from users about the conditions of their channels to decide which devices to serve through the multicast transmission and which via D2D communications. During this step, each device belonging to the MG sends to gNB its CQI values, concerning both its connection with gNB and the D2D links to its neighbors.
 
\subsection{Trustworthiness parameters collection}

The trustworthiness model proposed in this work foresees the setting of some parameters concerning the ``social state" of network nodes. During this step, cell nodes send to the gNB the information necessary to assess their trustworthiness in absence of direct interactions with the other nodes. Details on the model will be provided later, for now we point out that each device $i$ communicates the following trustworthiness parameters: its relationship factors (i.e., $F_{ij} \quad\forall j \textnormal{ friend of }i\textnormal{ in the cell}$), information to compute its centrality (i.e., $R_{ij}$), and information on its intelligence (i.e., $I_i$). The last term mainly refers to the computation capabilities of the device and it is better discussed in Section \ref{sec:trust_model}.
 
\subsection{Multicast and D2D configuration selection}

Thanks to the collected information, the gNB can plan: \textit{(i)} the set of registered UEs to serve through the multicast transmission via CMS; \textit{(ii)} the Modulation and Coding Scheme (MCS) to use for the multicast transmission in the CMS coverage area; \textit{(iii)} the set of registered UEs to serve via D2D communications because of their bad channel conditions; \textit{(iv)} the served UEs which can act as RNs by forwarding data towards the cell-edge users, thus forming the D2D pairs. 

Pseudocode shown in Table \ref{tab:d2d_selection} illustrates the sequence of steps executed to select the best multicast and D2D configuration. 

\begin{table}[htb]
\footnotesize
\centering
\caption{Multicast and D2D configuration selection in SeT-D2D.}
\rule{8.5cm}{0.3mm}
\begin{algorithmic}[1]
\State \textbf{Data}: $\mathnormal{U, CQI, CQI^c, CQI^d, SF, F, R, I, threshold}$
\State \textbf{Result}:  $\mathnormal{U^m, P^d, MDC}$
\ForAll{$c \in CQI$} 
\ForAll{$u \in U$}
\If{$CQI_u^c \leq c$}
\State Insert $u$ in $U^m$ 
\EndIf
\EndFor
\ForAll{$r \in U \setminus U^m$}
\State Find $s \in U^m | CQI_{rs}^d \neq 0$
\State Compute $st_{rs}(SF, F, R, I)$ \Comment{According to Eq. \ref{eq:st}}
\If {$st_{rs} \geq threshold$}
\State Insert s in $PR^r$
\Else
\State $s$ is not a possible transmitter for $r$
\EndIf
\ForAll{$p \in PR^r$}
\State Find $\max(CQI_{rp}^d)$
\State Update $P^d$
\State Insert $r$ in $U^d$
\EndFor
\EndFor 
\If {$U^m \cup U^d = U$}
\State Insert $MDC_c$ in $MDC$ \Comment{$MDC_c$ is related to CQI=c}
\EndIf
\EndFor
\State Find $m \in MDC | THR_m$ is max
\end{algorithmic}
\rule{8.5cm}{0.3mm}
\label{tab:d2d_selection}
\end{table}
Variables that are used are:

\begin{itemize}
\item \textit{U}, the set of registered UEs;
\item \textit{CQI}, the set of all CQIs (i.e., $[1,15]$);
\item $CQI^c$, the set of cellular CQIs of registered UEs (e.g., $CQI_u^c$ is the cellular CQI for UE \textit{u});
\item $CQI^d$, the set of D2D CQIs between nearby UEs in the cell (e.g., $CQI_{rs}^d$ is the D2D CQI between UEs \textit{r} and \textit{s});
\item \textit{SF}, the set of satisfaction factors $sf_{ij}$ related to the D2D pairs that interacted;
\item \textit{F}, the set of $F_{ij}$ related to pairs of UEs for whom there is some kind of social relationship;
\item \textit{R}, the set of $R_{ij}$ of registered UEs;
\item \textit{I}, the set of $I_i$ of registered UEs;
\item \textit{threshold}, is established to assess the eligibility of possible D2D pairs;
\item $U^m$, the set of registered UEs to serve in multicast;
\item $P^d$, the set of D2D pairs;
\item \textit{MDC}, the set of eligible multicast and D2D configurations;
\item $U \setminus U^m$, the set resulting from the difference between the sets \textit{U} and $U^m$;
\item $st_{rs}$, the service trust value between UEs \textit{r} and \textit{s};
\item $PR^r$, the set of possible D2D relays for user \textit{r};
\item $U^d$, the set of users that can be served in D2D as a transmitter is found for them;
\item \textit{THR}, representing throughput (e.g., $THR_m$ is the throughput related to configuration \textit{m}).
\end{itemize}

The multicast and D2D configuration selection is accomplished through an iterative procedure. First, the gNB sorts the received cellular CQI values from the lowest to the highest. Then, it analyzes every possible CQI value and, for each, determines  the subset of UEs which can decode the data transmitted with the correspondent MCS and the subset of UEs which, differently, are in worst channel conditions and must be served through D2D communications (lines 3-8). The selection of the most suitable D2D transmitters is the heart of our proposal. The gNB has stored information about the D2D CQIs of cell nodes; thus, it knows what the potential RNs for each D2D receiver could be. For each of them, it computes the value of \textit{service trust} to have a measure of their trust in the ability in providing services (lines 9-11), as will be detailed in Section \ref{sec:trust_model}. At this point, it classifies as ``not eligible'' the devices for which the service trust is not at least equal to an established threshold (lines 12-16). Among the remaining ones, the gNB selects, for each D2D receiver, the device with the highest D2D CQI towards it, i.e. the most efficient one (lines 17-21). As a consequence, D2D pairs are formed. This procedure is performed for each analyzed CQI value, in order to investigate all possible configurations. The term ``configuration" indicates the set of nodes to be served in multicast and those to be served in D2D, which depends on the MCS selected for the CMS transmission. Among all the eligible configurations (i.e., those in which all UEs, belonging to the MG, are able to receive data), the gNB finally selects the one that guarantees the maximization of transmission performance (lines 23-27). 

Once the selection of multicast and D2D configuration is accomplished, the operations aimed at securing D2D communications are carried out.

\subsection{D2D initialization}

The D2D receiver (in the following referred as $UE_i$), which wants to receive data, sends an initialization message to the gNB to communicate its identity (i.e., SUCI) and the key generated for the implementation of the DHKE algorithm. In this message and also in those to follow, the use of message authentication (i.e., HMAC) is envisioned for the integrity and authentication of transmitted data.

\subsection{D2D pair announcement}

After receiving the initialization message, the gNB verifies that the identity of the requesting UE is among those previously registered to the network. In positive case, the gNB communicates to each device of the D2D pair the identity of the other peer. Furthermore, it sends to the relay node, $UE_j$, the key previously received by $UE_i$, that it will need for the generation of the encryption key through the DHKE algorithm.

\subsection{Data transmission}

The gNB must sign data, before starting the multicast transmission to cell nodes, in order to attest their origin.
The gNB computes the signature by applying a public hash function (\textit{H}) to the data to be transmitted (\textit{D}) and by encrypting the resulting digest with its private key (\textit{pk}): $\sigma_1=H_{pk_1}(D)$. After that, it sends data to users with the best channel conditions, using CMS. $UE_j$, which has received data sent by the gNB, has to forward it to the previously notified D2D receiver; therefore, it carries out all the operations aimed at securing the D2D communication. It selects the key to be sent to the gNB so that $UE_i$ can generate, via the DHKE algorithm, the secret key used to encrypt data. After that, the RN itself generates the secret key and encrypts data. Eventually, $UE_j$ signs encrypted data and sends them to $UE_i$: $\sigma_j=H_{pk_j}(D')$, where $pk_j$ is the private key of $UE_j$ and \textit{D'} is the encrypted data.

\subsection{Data check}

After data reception from the RN, $UE_i$ must ensure that data comes from the transmitter previously announced by the gNB. If so, in order to accomplish the DHKE procedure and to obtain the plaintext, it requires to the gNB the key previously selected by $UE_j$. 
It is worth mentioning that, in the traditional DHKE algorithm, non-secret keys are exchanged directly between the communicating peers. Instead, in SeT-D2D the key exchange is mediated by the gNB, which represents a trusted third party. This helps in avoiding man-in-the-middle attacks that represent the main vulnerability of DHKE.

Data are accepted by $UE_i$ if and only if their origin is verified through the gNB's signature check (if this signature is not valid, then data may have been tampered by $UE_j$). In any case, $UE_i$ must send to the gNB a report in order to communicate some parameters on the quality of the transmission and, possibly, to report the fabrication of the original data, thus allowing the gNB to identify the attacker. The gNB waits for the report for an established waiting period. If it receives a report which announces a data security breach, then it checks by its own the information reported by $UE_i$. If the data breach is actually occurred and if the gNB ensures that data were sent by $UE_j$, then the gNB sets to $0$ the \textit{good transmission flag ($gtf_{ij}^l$)} related to the $l$ transmission from $UE_j$ to $UE_i$, in order to indicate that the D2D communication has not been successful. The same thing happens if the gNB does not receive any report by the end of the waiting period. On the contrary, if the report received from $UE_i$ indicates a well completed D2D transmission, the gNB sets to $1$ the $gtf_{ij}^l$. 

\begin{figure}[ht]
\centering
\includegraphics[scale=0.4]{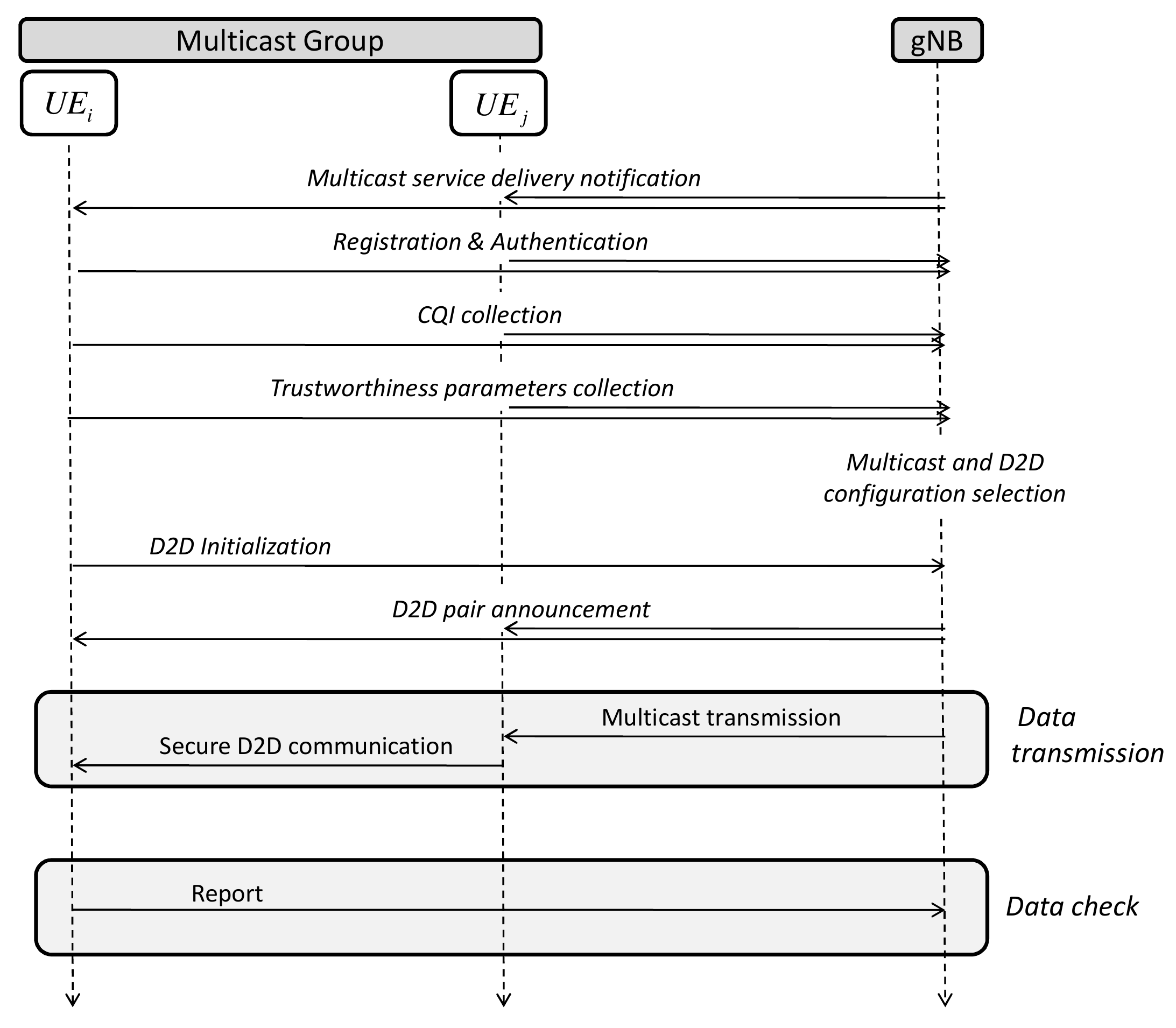}
\caption{The SeT-D2D protocol.} 
\label{fig:procedure}
\end{figure}

All steps of the proposed protocol are depicted in Fig. \ref{fig:procedure}.

\section{Trustworthiness Model}
\label{sec:trust_model}

In this section, we will describe the trustworthiness model we designed to allow the selection of trustworthy nodes acting as relays, and to establish secure D2D communications.
In the scenario under analysis, the gNB operates as a central controller that is able to assess whether the RN has transmitted data correctly or not (i.e., if it exhibits a malicious behavior or not).
The main problem in such a scenario is that, especially in the early stages, there is no information on the trustworthiness of the service provider (i.e., the RN). In fact, a requester node that never interacted directly with a RN is not able to make a judgment on it. Moreover, before being able to understand  the true nature of the RN (if it is a malicious or benevolent node), several interactions will have to take place.
In short, the system is affected by the cold start problem, which we cope to by designing a \textit{trustworthiness model} including not only a \textit{direct contribution}, accounting for the direct interactions between nodes, but also an \textit{indirect contribution}, accounting for structural and social properties independent of the transactions (interactions) between the nodes. For this reason, even in the absence of previous transactions, it is already possible to evaluate the maliciousness of the provider and, consequently, to reduce the cold start problem.

In the trustworthiness realm, with the term \textit{functional trust} the ability of a node to provide services \cite{Jos1} is meant. In this paper, in adopting the nomenclature used in \cite{Can}, we refer to the functional trust with the term \textit{service trust} for a more immediate understanding of the concept that it implies. In turn, service trust can be divided into \textit{direct service trust} and \textit{indirect service trust} \cite{Jos1}.

The direct service trust is measured through the opinion of the requester that directly interacted  with the provider in the past and, thus, can judge its ability in providing the requested service.
The indirect service trust results from the evaluation of the service that the RN has indirectly provided, i.e., through the opinions of the nodes that have already directly interacted with the service provider. Since in our scenario the gNB works as a trusted third party, the indirect service trust (and, more specifically, the service reputation) results from the real behaviour of the provider $j$ in all the past transactions in which $j$ has provided a service.
In a general trustworthiness model, also the concept of referral trust (named recommendation trust in \cite{Can}), that is the trust in the ability of a node to provide recommendations, can be included. In our scenario, referral trust is not analyzed since the centralized nature of the network architecture makes it lose meaning. For more details on the referral trust the reader can refer to \cite{Can,Jos1}.

In order to model the service trust, we must first introduce some factors, described by using concepts of graph theory. With $i$ and $j$ we represent two nodes of the graph and with $l$ the interaction (transaction) that occurs between them.

The first step is to find a way to evaluate the interaction between the specific pair of nodes $i$, $j$. Generally, in the literature, there are three parameters necessary to achieve this goal.
The first factor is $sf_{ij}^{l}$ that represents the \textit{satisfaction} of $i$'s $l^{th}$ interaction with $j$. This contribution allows a node $i$ to provide an evaluation of the service it has received by the provider $j$ \cite{Can,Chen,Peertrust,Subj,Obj}. 
Since, in our case, the gNB is a central entity able to assess whether the RN has transmitted data correctly or not, we consider:

\begin{equation}
\label{eq:sf}
sf_{ij}^{l} = gtf_{ij}^{l}
\end{equation}

where $gtf_{ij}^{l}$ is the good transmission flag for the specific interaction $l$ between node $i$ and node $j$. We recall that $gtf_{ij}^{l}$ assumes a value of 1 if the interaction between $i$ and $j$ is successful, 0 otherwise.

Generally, the satisfaction $sf_{ij}^{l}$ can also keep into account other factors in addition to the contribution of security, such as throughput $TP_{ij}^{l}$ and delay $D_{ij}^{l}$. In such a case, the expression of $sf_{ij}^{l}$ is the following:

\begin{equation}
sf_{ij}^{l} = \chi gtf_{ij}^{l} + \psi TP_{ij}^{l} + \sigma D_{ij}^{l}
\end{equation}

We decided to consider Eq. \ref{eq:sf} for computing the satisfaction of the interaction to specifically focus on security.
In fact, by assigning the good transmission flag calculated by the gNB as the satisfaction value, we have a correct and objective evaluation of the occurred data transmission.

The second parameter needed to evaluate the transaction between the specific pair of nodes $i$ and $j$ is $s\omega_{ij}^{l}$, that represents the \textit{importance} (or relevance) of $i$'s $l^{th}$ interaction with $j$. 
It is used to discriminate important transactions from irrelevant ones \cite{Can,Chen,Peertrust,Subj,Obj}.

The last term to consider for evaluating the interaction is $s\delta_{ij}^{l}$, that represents the \textit{decaying factor} of $i$'s $l^{th}$ interaction with $j$ \cite{Can,Chen}. Looking at Eq. \ref{eq: quality utility}, with $sh_{ij}$ we mean the size of $i$'s service (interaction) history with $j$, that is the total number of interactions occurred between $i$ and $j$ \cite{Can}, while $l$ represents the number of the current interaction between $i$ and $j$. In particular, the decaying factor is expressed as:

\begin{equation}
\setlength{\nulldelimiterspace}{0pt}
s\delta_{ij}^{l}=\left\{\begin{IEEEeqnarraybox}[\relax][c]{c's} \mu{\frac{l}{sh_{ij}}} + \nu{\frac{1}{ln(|t-t^I|)}}, & for ${ln(|t-t^I|)} > 1$ \\ \mu{\frac{l}{sh_{ij}}} + \nu, & otherwise
\end{IEEEeqnarraybox}\right.
\label{eq: quality utility}
\end{equation}

where $t$ is the actual time and $t^I$ is the occurrence time (generation time) of this interaction.
We must also impose the existence condition of the logarithm, that is $|t-t^I| > 0$.
Variable $\mu$ and $\nu$ represent the weights of the two contributes and must be set on the basis of the scenario under investigation.
In particular, the first contribution of the decaying factor takes into account the number of interactions occurred after the one under analysis (current validity of the interaction), while the second contribution accounts for the time elapsed from the considered interaction (recency of the interaction). The first term is derived from \cite{Can}, while the second from \cite{Chen}.
In particular, the first contribute is important in scenarios in which many interactions occur in a short time interval. In fact, even if an interaction took place recently, since other interactions occurred later, it may be not consistent with the current situation. The second contribution has been added because it can allow to account also for cases where not many interactions occur. In these scenarios, the interaction considered may be the last occurred between that specific pair of nodes, but the same can be no longer trustworthy if it happened a long time ago. By merging these two contributions, as explained above, we expect to obtain a decaying factor that can more accurately consider both the decay phenomena.

So far we have described the parameters that underlie the calculation of the direct contribution of service trust. As previously discussed, one of the main benefits of the trustworthiness model is that it allows to evaluate the trustworthiness of a node even in the absence of direct transactions, by means of the parameters contained in the indirect contribution. 

The first term included in the indirect contribution of the service trust is $F_{ij}$, that is the \textit{relationship factor} indicating the type of relation that connects $i$ with $j$ \cite{Subj,Obj}. It represents a unique characteristic of the SIoT. We briefly describe the types of existing social relationships \cite{Iera1}. The Owner Object Relationships (OORs) are established  between two objects that belong to the same owner. In this kind of relation, it is very unlikely to find a malicious node. The Co-Location Object Relationships (C-LORs) connect domestic objects, the Co-Working Object Relationships (C-WORs) link objects of the same workplace. The Social Object Relationships (SORs) are established between objects that meet occasionally. The Parental Object Relationships (PORs) are created between objects of the same model. The process of establishment of these relationships precedes the interactions, as it is mainly based on the number and duration of previous contacts  occurring between the devices. We used SWIM simulator \cite{SWIM} to generate traces of people's mobility. It has been properly modified to obtain traces of the mobility of devices owned by people.
A trustworthiness value has been associated with each social relation as shown in Table \ref{tab:soc_rel}, according to \cite{Subj}.

\begin{table}[htbp]
\caption{Social relationships trustworthiness possible values.}
\centering
\begin{tabular}{p{1.8cm}p{4.5cm}p{1.2cm}}
\hline
\textbf{Relationship} & \textbf{Description}& \textbf{Trust} \\
\hline
OOR & Objects owned by the same person & 0.9\\
C-LOR & Objects sharing experiences & 0.8\\
C-WOR & Objects sharing public experiences & 0.8\\
SOR & Objects in contact for owner's relations & 0.6\\
POR & Objects with production relations & 0.5\\
No relationship & & 0.1 \\
\hline
\end{tabular}
\label{tab:soc_rel}
\end{table}

The second term of the indirect contribution is also of a social nature and is $ R_{ij}=\frac{|K_{ij}|}{|N_{i}| } $, that is the \textit{centrality} of $j$ in the life of $i$, where $|K_{ij}|$ represents the \textit{common friends} between $i$ and $j$, and $|N_{i}|$ is the \textit{neighborhood} of node $i$ \cite{Subj,Obj}. This term is very important because if a node has many relationships, it is expected to assume a central role in the network in terms of leadership, efficiency in problem solving, and personal satisfaction of participants. Furthermore, if two nodes have many friends in common, it is likely that they have similar evaluation parameters about building relationships.

The third parameter of the indirect contribution of the service trust is $I_{j}$, that is the \textit{intelligence} of $j$, representing its computational capabilities \cite{Subj,Obj}. It is a static characteristic of the objects since it does not vary over the time, but depends only on the type of the object considered. We expect that a smart object has more capabilities to cheat with respect to a ``dummy'' object, and can lead to riskier transactions. 

As stated in Section \ref{sec:proposal}, $UE_i$ will send to the gNB the $F_{ij}$ and $R_{ij}$ values related to each of its neighbors.
Furthermore, it sends its own $I_i$ value.

The last term of the indirect contribution is \textit{service reputation} $sr_j$, that we will better explain at the end of the section. 

Now we introduce two direct contributions: the competence belief and the integrity belief.
The \textit{service competence belief} measures how well an acquaintance satisfied the needs of past interactions \cite{Can}.

\begin{equation}
scb_{ij}={\sum_{l=1}^{sh_{ij}}(sf_{ij}^{l} s\omega_{ij}^{l} s\delta_{ij}^{l}) \over \sum_{l=1}^{sh_{ij}}(s\omega_{ij}^{l} s\delta_{ij}^{l})}
\end{equation}

This term is fundamental because it stores the past history of the transactions occurred between each pair of nodes. In fact, we recall that $sh_{ij}$ is the size of $i$'s service (interaction) history with $j$, that is the total number of interactions occurred between $i$ and $j$, while $l$ represents the number of the current interaction between $i$ and $j$.

The \textit{service integrity belief} is the level of confidence in the predictability of future interactions \cite{Can}.

\begin{equation}
sib_{ij}=\sqrt{{1\over{sh_{ij}}} \sum_{1}^{sh_{ij}} (SO_{ij}^{rec}-SO_{ij}^{lon})^2}
\label{eq:sib}
\end{equation}

Small values of integrity translate into a more predictable behavior of $j$ in future interactions.
The idea is to consider not only the degree to which a node has satisfied past interactions, but also the deviation in the degree of satisfaction of the recent interactions with respect to the remote ones. The concept of predicting the degree of satisfaction of future interactions is found in literature in Bayesian Systems and Belief Theory models (including Subjective Logic \cite{Jos1}). These models use the Probability Density Function (PDF) and the Expected Values of PDFs. A less complex way to try to calculate predictability of future interactions is by using the standard deviation \cite{Can}. The novelty introduced in this model is to use the standard deviation for calculating the predictability of future interactions and to consider the service opinion long as the mean value. The advantage is that this method is simple and allows us greater control. Furthermore, it does not resort to complex formulas (like Bayesian Theory and Belief Theory), which also include other parameters such as uncertainty.

The two terms we use to calculate service integrity belief are \textit{service opinion long} and \textit{service opinion recent} as in \cite{Subj,Obj}. They can be expressed as:

\begin{equation}
SO_{ij}^{lon}={\sum_{l=1}^{L^{lon}}(sf_{ij}^{l} s\omega_{ij}^{l} s\delta_{ij}^{l}) \over \sum_{l=1}^{L^{lon}}(s\omega_{ij}^{l} s\delta_{ij}^{l})}
\end{equation}

\begin{equation}
SO_{ij}^{rec}={\sum_{l=1}^{L^{rec}}(sf_{ij}^{l} s\omega_{ij}^{l} s\delta_{ij}^{l}) \over \sum_{l=1}^{L^{rec}}(s\omega_{ij}^{l} s\delta_{ij}^{l})}
\end{equation}

They represent the long-term and the short-term service opinion of $i$ about $j$ and they are based on the satisfaction of $i$ with respect to the services provided by $j$.
${L^{lon}}$ represents the long-term opinion temporal window and ${L^{rec}}$ the short-term opinion, with ${L^{lon}}>{L^{rec}}$ and $l$ indexes from the latest to the oldest transaction.

Now we are ready to calculate the service trust between nodes \textit{i} and \textit{j} ($st_{ij}$):

\begin{equation}
\label{eq:st}
\begin{multlined}
st_{ij}=\Biggl( {log(sh_{ij}+1)\over{1+log(sh_{ij}+1)}}\Biggr)(\beta_{1} scb_{ij}-\beta_{2} sib_{ij}) +\\
+\Biggr({1\over{1+log(sh_{ij}+1)}}\Biggr)(\gamma sr_{j} + \epsilon F_{ij} + \zeta R_{ij} + \theta (1-I_{j}))
\end{multlined}
\end{equation}

In particular, we have used the following structure in the previous formula:

$ \alpha * Direct Experience + \beta * Indirect Experience $ 

where $\alpha$ grows and $\beta$ decreases with the number of interactions. Values of $\alpha$ and $\beta$ are taken from the literature \cite{Subj}.

Thanks to this structure, as the number of interactions increases, we will assign an increasing weight to the direct experience contribution. A similar structure is used in \cite{Can,Chen,Fuzzy,Scalable,Subj,Obj}.
In the direct experience contribution, the first term (service competence belief) corresponds to the direct functional trust; while in the indirect experience contribution, the first term ($sr_{j}$ service reputation) corresponds to the indirect functional trust \cite{Jos1}. In order to ensure that $st_{j}$ assumes values between 0 and 1, the sum of the constants of both the direct and indirect contributions will be equal to 1. 

As previously discussed, in the contribution of indirect experience, we also consider parameters, such as relationship factor, centrality, and intelligence, that allow us to calculate the service trust even in the absence of direct interactions \cite{Subj,Obj}, in order to partially solve the cold start problem.

The last term constituting the indirect contribution of the service trust  is the service reputation $sr_ {j}$. Generally, a node can get information about the ability in providing service of $j$ by asking the opinion to other nodes that have already received a service from $j$. In particular, in distributed scenarios, service reputation can be calculated by considering only the subjective opinion of a specific subset of nodes to which $j$ has already provided a service. Often, this subset is the neighborhood of the requester node $i$ since there is no trusted third party.

In our scenario, the gNB works as a central controller and computes the service reputation by considering the real behaviour of the provider $j$ as:

\begin{equation}
sr_{j}={1\over{n_{j}}}{\sum_{g=1}^{|N_{j}|}(sf_{gj})} 
\label{eq:sr}
\end{equation}

where $|N_{j}|$ is the set of nodes that have already interacted with $j$ and $n_{j}$ is the total number of transactions in which $j$ has supplied a service. 

As can be seen from the structure of the designed model, we do not decide to insert a threshold to definitively exclude the nodes that did not behave trustworthy, for several reasons.
First of all, in any case the model will tend to naturally assign low values of service trust to malicious nodes, thus excluding them (not definitively) even without using a threshold. A second reason is that, unlike threshold-based models, our model is able to recover the benevolent nodes which behaved badly for a few interactions (for example, because they are temporarily infected by viruses). Obviously, the recovery process is not immediate and depends on the past history of the node (past interactions of the node). The increase in service trust is gradual and dependent on the behavior of the node. 
We conclude by showing in Table \ref{tab:param} the weights of the parameters used to carry out the simulations. These values are relaying on existing works \cite{Can,Subj,Obj}.

\begin{table}[htbp]
\caption{Values of the weights of parameters.}
\centering
\begin{tabular}{p{2.0cm}p{3.0cm}p{1.5cm}}
\hline
\textbf{Weight} & \textbf{Parameter} & \textbf{Value} \\
\hline
$\beta_{1}$ & Competence & 1\\
$\beta_{2}$ & Integrity & 0.5\\
$\gamma$ & Reputation & 0.5\\
$\epsilon$ & Relationship Factor & 0.175\\
$\zeta$ & Centrality & 0.175\\
$\theta$ & Intelligence & 0.15\\
\hline
\end{tabular}
\label{tab:param}
\end{table}

\section{Performance Evaluation}
\label{sec:5}

We tested the performance of the proposed protocol via the Matlab tool. 

The considered scenario consists of 100 devices uniformly distributed in a 100 m x 100 m cell. Inside the multicast group, including all terminals, a portion of devices is served according to a CMS approach, while those in worst channel conditions receive data via D2D connections. 
A New Radio (NR) frame with transmission numerology $\mu = 0$ is considered in the simulations, that is a frame with 15 KHz subcarrier spacing composed by 10 slots, each lasting 1 ms. A bandwidth of 20 MHz with 100 RBs is available. 
The NR frame used in the simulations consists of six slots in format 0, three slots in format 1, and one slot in format 2, organized as in a TDD LTE frame type 2 configuration 3.  
The Inband D2D mode is chosen, therefore uplink slots are reserved to D2D communications. In downlink slots, the multicast transmission takes place.

The following metrics are used to assess the performance of the proposed protocol:
\begin{itemize}
\item \textit{Percentage of wasted capacity} on the D2D link caused by the selection of untrustworthy transmitters.
\item \textit{Mean number of non-corrupted received kbits}, which indicates the amount of data correctly downloaded in D2D, as transmitted by non-malicious relays.
\item \textit{Percentage of malicious relay selection}, computed as the percentage of frames, over all simulation time, in which at least one malicious relay has been selected.
\end{itemize}

\begin{figure}[h]
\centering
{\includegraphics[scale=0.45]{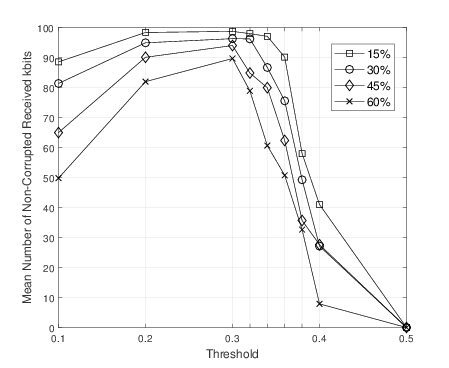}}
\caption{Mean number of non-corrupted received kbits under varying maliciousness threshold.}
\label{fig:THR}
\end{figure}

In the following, before discussing the performance offered by the proposed mechanism, we briefly demonstrate how some features introduced in our trustworthiness model can help in overcoming the drawbacks pointed out in Section \ref{sec:rw}.
Then, we will show, the performance enhancement that the proposed SeT-D2D protocol can provide with respect to: \textit{(i)} exploiting only information derived from past interactions, thus regarding security (Se-D2D), and \textit{(ii)} a security- and trust-unaware algorithm (D2D).
Note that the Se-D2D comparison protocol corresponds to those defined in \cite{seds}, while the D2D protocol represents a legacy D2D communication.
In the remainder of the Section, we analyze in details how the proposed approach works under two different kinds of attack model. In case of \textit{on-off attacks} (inspired by \cite{onoff}), malicious nodes exhibit their malevolent nature only during on periods. Differently, \textit{receiver-selective attacks} reflect the case of a node that behaves maliciously only toward specific receivers.

As stated in Sec. \ref{sec:trust_model}, only nodes with an $st$ value higher than a predetermined threshold are considered as possible relay nodes. Fig. \ref{fig:THR} shows the results of an experimental analysis we carried out with the aim to define the most appropriate value of such a threshold. Since, under all analyzed percentages of malicious nodes (ranging from 15\% to 60\%), the mean number of non-corrupted received kbits reaches the highest value around the threshold value of 0.3, in all simulations results that will be shown in the following the threshold is set to this value. As expected, Fig. \ref{fig:THR} also allows to appreciate that higher amount of non-corrupted data is delivered to receiver nodes under lower percentage of malicious nodes.

\subsection{\change{Proof of concept of the trustworthiness model}}

\begin{figure}[h]
\centering
{\includegraphics[scale=0.23]{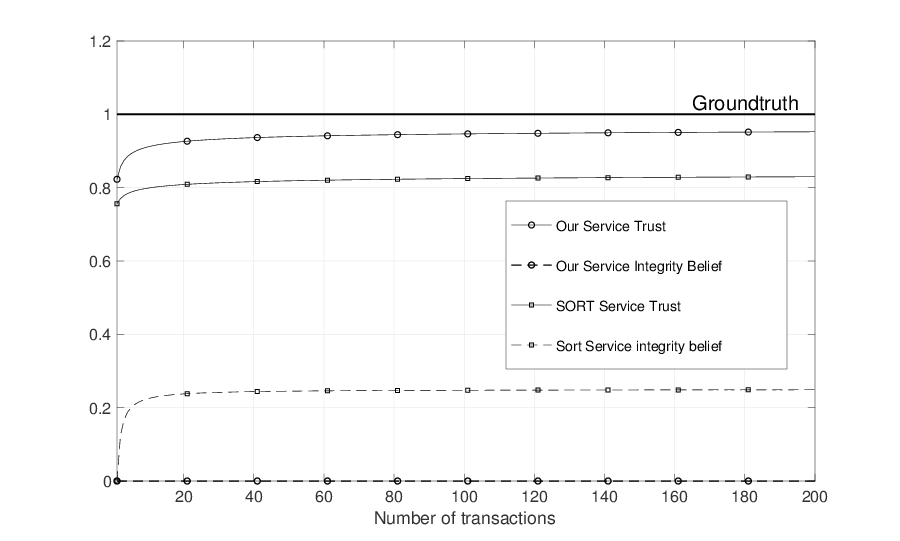}}
\caption{Performance comparison obtained by our model with different service integrity belief formulas, in a scenario with $sf$ (service satisfaction) always equal to 1.}
\label{fig:CI}
\end{figure}

We first focus on the calculation of the service integrity belief $sib_{ij}$ measured by node $i$ by referring to the service received by provider $j$ (see Eq. \ref{eq:sib}). 
Similarly to \cite{Can}, we follow an approach based on the standard deviation since it is simpler than the Bayesian and Belief models and allows greater control. 
In Fig. \ref{fig:CI}, we compare our approach in case service integrity belief is calculated according to Eq. \ref{eq:sib} or with the SORT integrity formula proposed in \cite{Can}. We analyze the ideal case in which node $j$ always provides an excellent service (i.e., $sf$ is always equal to 1 and corresponds to the Ground truth). In such a condition, since the deviation in the degree of satisfaction of recent interactions with respect to the remote ones is null, node $i$ expects that $j$ exhibits the same behavior shown in the past.
By looking at Fig. \ref{fig:CI}, we can appreciate that, by using SORT, we tend to overestimate the service integrity belief and underestimate the service trust. Differently, the adoption of our model allows to obtain the expected service integrity belief (the curve is always equal to 0) and a service trust very close to the Ground truth.
Similar conclusions can be drawn also in non-ideal scenarios (i.e., when node $j$ does not always provide satisfactory services).
 
\begin{figure}[t]
\centering
\subfigure[]{\includegraphics[scale=0.23]{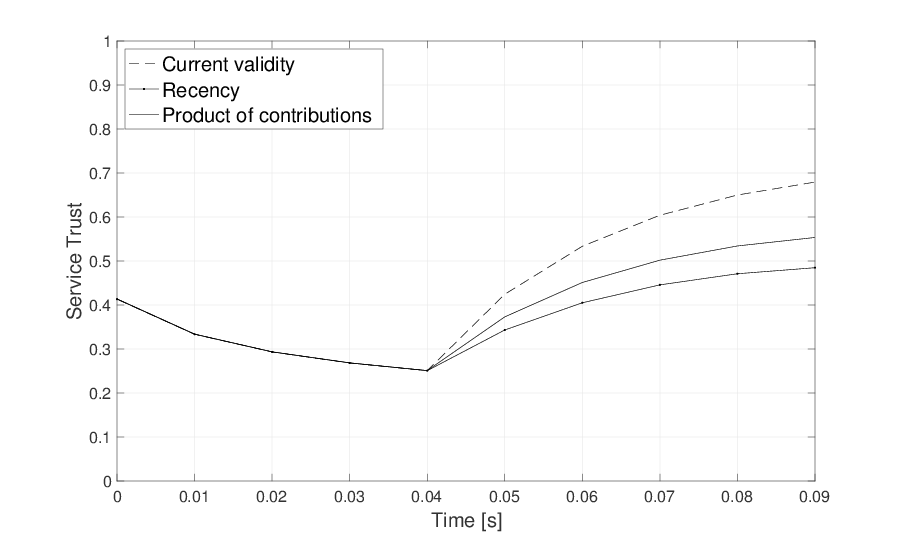}}
\label{fig:Short}
\subfigure[]{\includegraphics[scale=0.23]{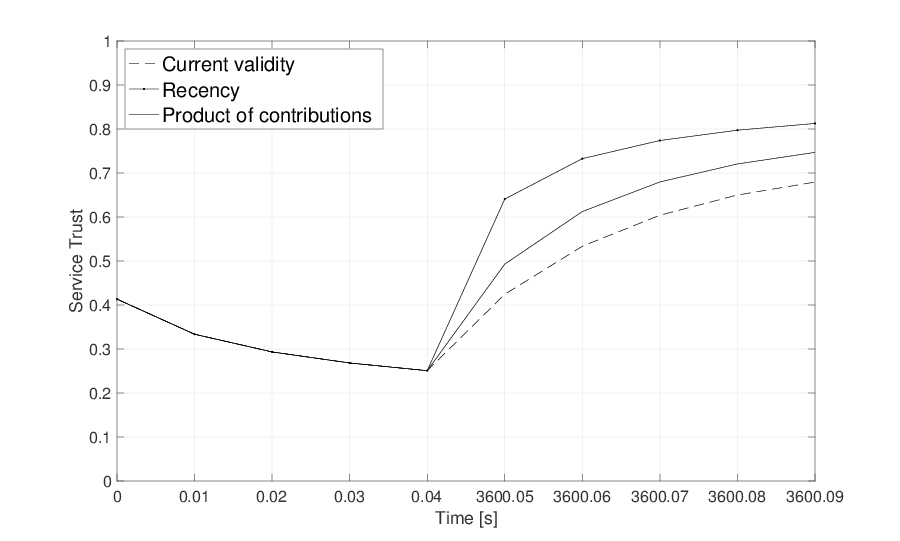}}
\label{fig:Long}
\caption{(a) Short time interval scenario, and (b) Long time interval scenario.}
\label{fig:SL}
\end{figure}

Let us focus on the decaying factor ${s\delta}^l_{ij}$ of the $l-th$ transaction between node $i$ and provider $j$ (see Eq. \ref{eq: quality utility}). We remark that, in the literature, some models (such as \cite{Can}) consider only the \textit{cardinal contribution} that keeps into account the number of transactions occurred after the considered one (current validity of the interaction), while other research works (such as \cite{Chen}) focus only on the \textit{temporal contribution} related to the time elapsed from the $l-th$ transaction (recency of the interaction). 
With the aim to provide a good approximation of the decaying factor, our model considers both contributions. 
We analyze two scenarios in which 10 transactions take place either in a short time interval (Fig. \ref{fig:SL}(a)) or in a long time interval (Fig. \ref{fig:SL}(b)), respectively. 
In the ``short time interval'' scenario, a transaction takes place every 0.01 s. Differently, in the ``long time interval'' scenario there is a pause of 3600 s between the fifth and the sixth transaction. In both cases, we assume that only the last five transactions have $sf = 1$ (i.e., node $i$ evaluates the service received by $j$ in the last five transactions satisfactory).
Fig. \ref{fig:SL}(a) shows that, when considering only the \textit{cardinal contribution} ($\mu$ = 1, $\nu$ = 0), the weight of last 5 transactions increases compared to the case in which only the \textit{temporal contribution} is taken into account ($\mu$ = 0, $\nu$ = 1). 
The third curve, lying in the middle, represents the case in which both contributions are considered and weights $\mu$ and $\nu$ are set, as an example, to 0.5. 
Differently, by looking at Fig. \ref{fig:SL}(b), we can infer that the \textit{temporal contribution} alone ($\mu$ = 0, $\nu$ = 1) makes the last 5 transactions more relevant. 
The straightforward reason for this behaviour is that, in scenarios wherein  the considered transaction is recent in time, although followed by several further interactions (i.e. the situation represented by our ``short time interval'' scenario), models that keep into account only the \textit{temporal contribution} could reduce the decay factor since they evaluate only the recency of the interaction. However, the likely high number of interactions occurred after the considered one might have led to changes that the temporal contribution alone is unable to catch. Indeed, in this case accounting also for the \textit{cardinal contribution} would make our model more effective. 
Differently, in scenarios in which the considered interaction is not recent, although it is among the last ones occurred (i.e. the situation represented by our ``long time interval'' scenario), a higher weight given to the \textit{temporal contribution} could increase the effectiveness of the model.
We can conclude that both the decay contributions should be accounted for and that, in doing so, the proposed solution becomes flexible enough to allow getting all intermediate performance levels that lie in between the two limit cases represented above. This is achieved by properly setting the weigths $\mu$ and $\nu$ in the calculation of the decay factor.

As a final remark, it is worth stressing that our proposed trust model is explicitly designed for SIoT scenarios. Thanks to the presence of the intelligence, centrality, and relationship factor parameters, it allows the calculation of an indirect contribution even in the total absence of transactions (i.e. in the absence of service reputation), which makes it possible to mitigate the cold start problem. This will emerge in Fig.s \ref{fig:PERC}, \ref{fig:VARDIM} of Section \ref{sec:5}, from the comparison between Se-D2D and SeT-D2D. Besides, it gives weights to parameters so as to allow a greater adaptability to the considered scenarios. This last advantage emerges in  Fig. \ref{fig:var_beta} of Section \ref{sec:5}. 

\subsection{\change{Proof of concept of the SeT-D2D protocol}}

\begin{figure*}[t]
\centering
\subfigure[]{\includegraphics[scale=0.358]{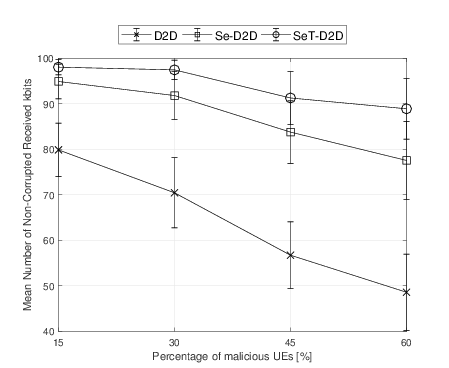}}
\label{fig:PERC_BYTES}
\subfigure[]{\includegraphics[scale=0.358]{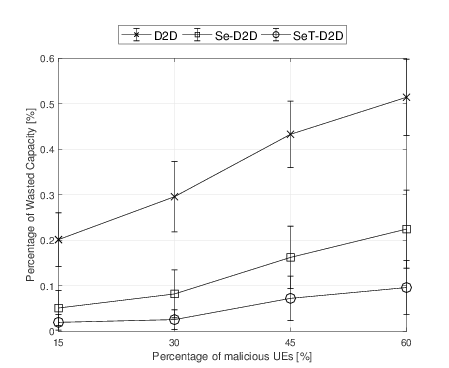}}
\label{fig:PERC_WASTE}
\subfigure[]{\includegraphics[scale=0.358]{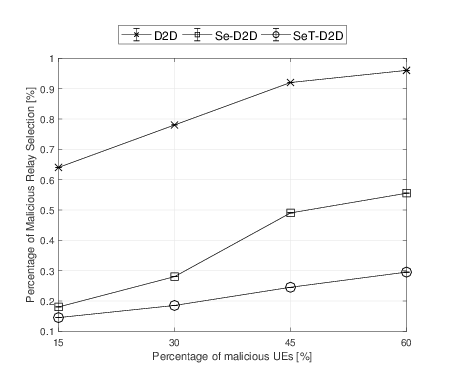}}
\label{fig:THR_SIM}
\caption{(a) Mean number of non-corrupted received kbits, (b) percentage of wasted capacity, and (c) percentage of malicious relay selection under varying percentage of malicious users per cell.}
\label{fig:PERC}
\end{figure*}

Fig. \ref{fig:PERC} shows the benefits that the proposed SeT-D2D can bring with respect to Se-D2D and D2D in terms of mean number of non-corrupted received kbits (Fig. \ref{fig:PERC}(a)), percentage of wasted capacity (Fig. \ref{fig:PERC}(b)) and percentage of malicious relay selection (Fig. \ref{fig:PERC}(c)) for an increasing percentage of malicious nodes. Plots show that utilizing information deriving from security is of primary importance and that a further considerable performance enhancement can be achieved when social trustworthiness is also kept into account. In fact, SeT-D2D protocol is able to deliver the highest amount of non-corrupted kbits, thus leading to the lowest percentage of wasted capacity, thanks to the infrequent selection of malicious relay nodes. 

\begin{figure}[t]
\centering
\subfigure[]{\includegraphics[scale=0.4]{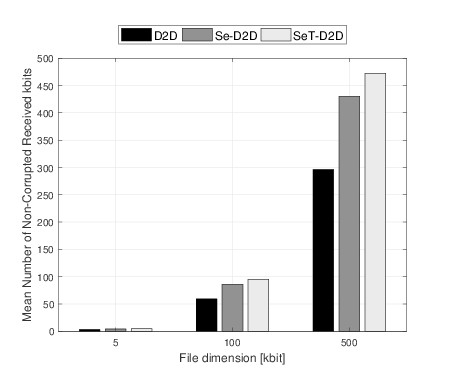}}
\label{fig:VARDIM1}
\subfigure[]{\includegraphics[scale=0.4]{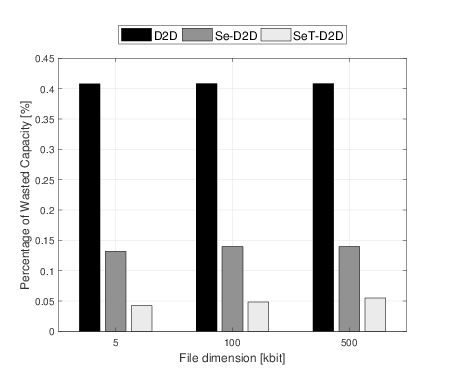}}
\label{fig:VARDIM2}
\caption{(a) Mean number of non-corrupted received kbits, and (b) percentage of wasted capacity under increasing file dimension.}
\label{fig:VARDIM}
\end{figure}

To further prove the effectiveness of the proposed SeT-D2D protocol, we present in Fig. \ref{fig:VARDIM} an analysis under increasing file dimension, representative of the performance that our proposal could assure in different use cases, ranging from alert messaging to file downloading. Results testify that, for all considered file dimensions, SeT-D2D exhibits the best performance in terms of both mean number of non-corrupted received kbits (Fig. \ref{fig:VARDIM}(a)) and percentage of wasted capacity (Fig. \ref{fig:VARDIM}(b)). We also point out that this last metric is insensitive to file dimension while the former shows an increasing trend.   

\subsection{Analysis of on-off attacks}

A malicious D2D transmitter could follow different attack models. 
In the following analysis, the response of the trustworthiness model presented in our work is shown under three different \textit{attack rates} (i.e., 30\%, 50\%, and 80\%), indicating the percentage of simulations over the total number of executed runs in which the relay acts maliciously against all its receivers. These are referred as \textit{on-off attacks}. 
Attack simulations can be consecutive or have an irregular pattern. 

\begin{figure*}[t]
\centering
\subfigure[]{\includegraphics[scale=0.34]{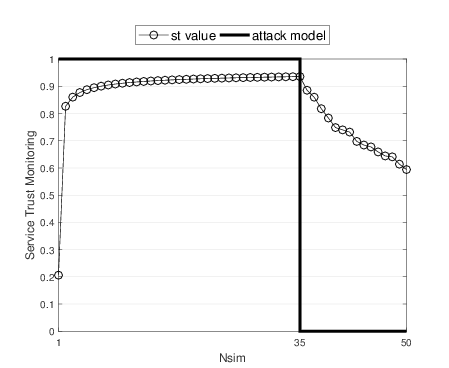}}
\label{fig:st_time_30start}
\subfigure[]{\includegraphics[scale=0.34]{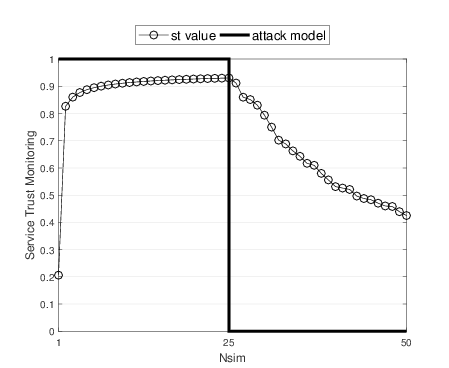}}
\label{fig:st_time_50start}
\subfigure[]{\includegraphics[scale=0.34]{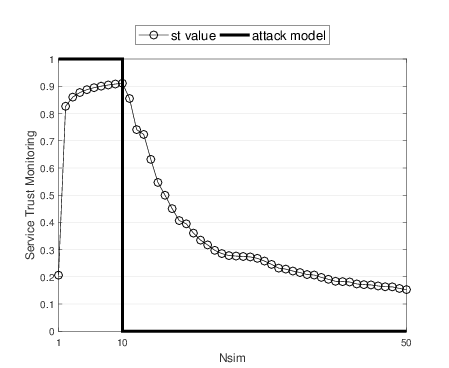}}
\label{fig:st_time_80start}
\subfigure[]{\includegraphics[scale=0.34]{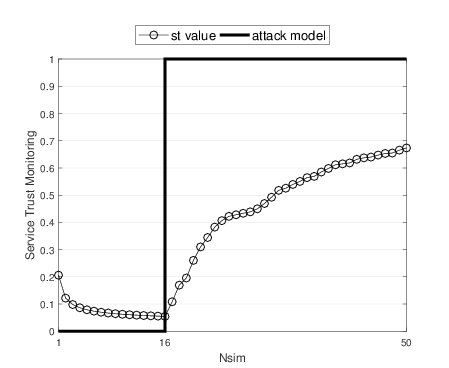}}
\label{fig:st_time_30end}
\subfigure[]{\includegraphics[scale=0.34]{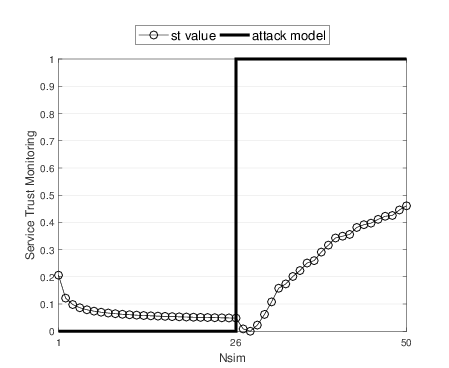}}
\label{fig:st_time_50end}
\subfigure[]{\includegraphics[scale=0.34]{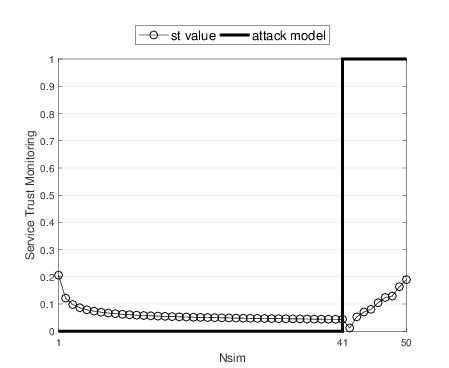}}
\label{fig:st_time_80end}
\caption{Service trust monitoring in case of on-off attack model with a final attacker activity equal to (a) 30\%, (b) 50\%, and (c) 80\%; vice versa, with an initial attacker activity equal to (d) 30\%, (e) 50\%, and (f) 80\%.}
\label{fig:onoff_atk}
\end{figure*}

In case of \textit{consecutive} attacks, the node could exhibit its malicious behavior after a period of time in which it hides its real nature by avoiding to carry out any attacks. Vice versa, it could behave maliciously initially, then moving to a good behavior. Fig.s \ref{fig:onoff_atk} show how $st_{ij}$ evolves over time for the D2D pair $ij$ that interacted during all simulations. The relay node behaves correctly when the attack model profile is equal to 1. Contrarily, it performs a malicious behaviour when the profile is equal to 0. Two complementary attack models are analyzed. In the first one, the malicious transmitter initially performs a good behavior and then attacks all its receivers for a number of simulations related to each attack rate (Fig.s \ref{fig:onoff_atk}(a), (b), and (c)). The opposite attack profile is considered in Fig.s \ref{fig:onoff_atk}(d), (e), and (f). All graphs show a trend in the value of the service trust that is consistent with the evolution of the attack model, as the value of $st_{ij}$ grows as long as the relay transmits data correctly, and begins to decrease when it exhibits its malicious nature. This proves that the proposed trustworthiness model produces a service trust value that reflects the nature of the node, thus representing an effective estimate of node's trustworthiness.

\begin{figure}[h]
\centering
{\includegraphics[scale=0.4]{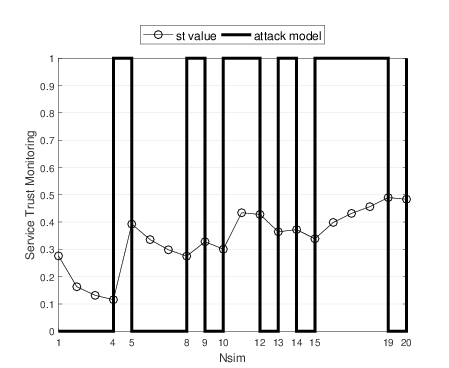}}
\caption{Service trust monitoring in case of irregular attack model.}
\label{fig:irr_atk}
\end{figure}

Fig. \ref{fig:irr_atk} shows the trend of $st_{ij}$ under an \textit{irregular} attack model characterized by a discontinuous behavior of the malicious relay due to frequent transition from correct to corrupted D2D transmission. Also this figure demonstrates that the proposed trustworthiness model is able to produce a service trust value that is coherent with the attack model profile. 

\begin{figure}[h]
\centering
{\includegraphics[scale=0.4]{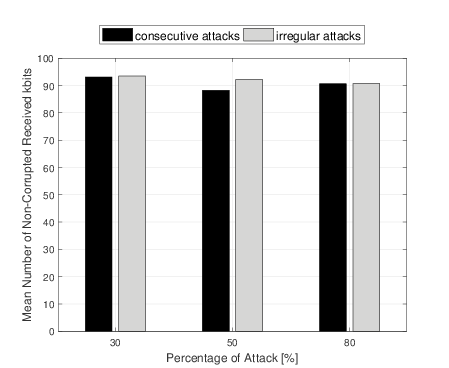}}
\caption{Mean number of non-corrupted received kbits under varying attack rate.}
\label{fig:cons_vs_irr}
\end{figure}

Fig. \ref{fig:cons_vs_irr} is a further proof that the trustworthiness model reacts well to both types of attack model (consecutive and irregular) since, by varying the attack rate, the amount of non-corrupted data delivered to the D2D receiver is similar in the two cases.

\begin{figure}[h]
\centering
{\includegraphics[scale=0.4]{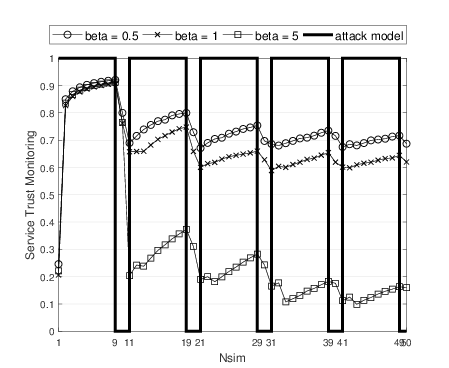}}
\caption{Service trust monitoring in case of periodic attack model under varying $\beta_2$ values.}
\label{fig:var_beta}
\end{figure}

A further test of the trustworthiness model responsiveness to on-off attacks is done by simulating a \textit{periodic} attack model in which the malicious relay alternates a good and a bad behavior at regular intervals. Fig. \ref{fig:var_beta} depicts how $st_{ij}$ evolves when different $\beta_2$ values are used to weigh the integrity of the D2D interacting couple. 
As discussed in Sec. \ref{sec:trust_model}, when evaluating the direct experience in the interaction between receiver \textit{i} and malicious relay \textit{j}, a proper decrease in the value of $st_{ij}$ is guaranteed by an appropriate proportion between competence (i.e., $scb_{ij}$) and integrity (i.e., $sib_{ij}$) values of the D2D pair (Eq. \ref{eq:st}). The higher the value of $\beta_2$, the greater the decrease in $st_{ij}$. This may not be an advantage when a node manifests a single malicious behavior, not necessarily intentional, as its trust value falls considerably, making its recovery very slow. With this analysis we point out that the choice of the $\beta_2$ weight is very important and must be made based on the type of application and the severity to be attributed to incorrect behaviors. In all other simulations, $\beta_2$ is set to 0.5.

\subsection{Analysis of receiver-selective attacks}

Unlike on-off attacks, with \textit{receiver-selective attacks}, a malicious relay exhibits its bad nature at any time but not with all data recipients. Purpose of this analysis is to observe the effect of these attacks on the reputation of the malicious node and, consequently, on the evaluation of its service trust.

\begin{figure}[t]
\centering
\subfigure[]{\includegraphics[scale=0.4]{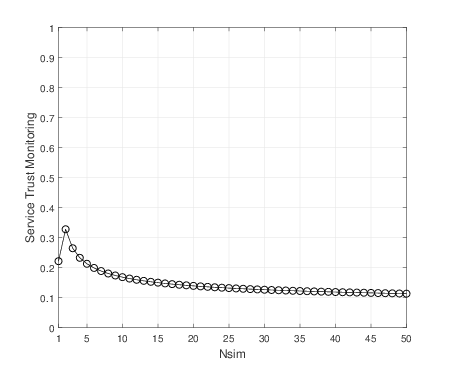}}
\label{fig:victim_mean}
\subfigure[]{\includegraphics[scale=0.4]{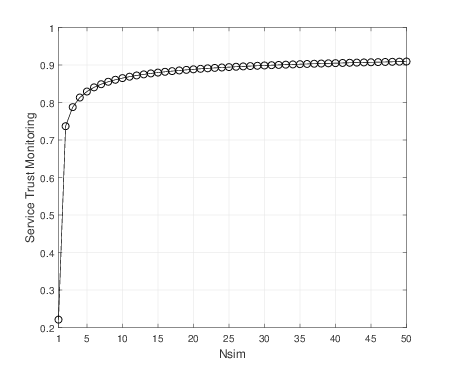}}
\label{fig:novictim_mean}
\caption{Service trust monitoring between malicious relay and: (a) victim node, (b) non-victim node.}
\label{fig:st_sr}
\end{figure}

The curves in Fig.s \ref{fig:st_sr} represent the service trust value referred to the interacting pairs in all executed simulations.
Fig. \ref{fig:st_sr}(a) shows the trend of $st_{ij}$ for the pair ${ij}$, where \textit{j} is the malicious relay and \textit{i} is its only victim receiver among the three receivers that it serves. Differently, in Fig. \ref{fig:st_sr}(b), $st_{wj}$ is computed between the malicious relay \textit{j} and the non-victim receiver \textit{w}. 
By looking at Fig.s \ref{fig:st_sr}, we can see that $st_{ij}$ has a trend which depends on the behavior that the relay exhibits with the peer. In fact, the gNB calculates for non-victim node (see Fig. \ref{fig:st_sr}(b)) a service trust value that is substantially higher than the one computed for the victim node (see Fig. \ref{fig:st_sr}(a)), due to the different direct contribution values. This proves that the higher the information available regarding security, the lower the influence of reputation on service trust and that our mechanism is able to fastly provide a differentiated estimation of the real malicious/non-malicious nature of the relay to victim and non-victim nodes, respectively. 

\subsection{\change{Security analysis}}

\change{This section is aimed at highlighting the security assurances that the SeT-D2D protocol offers. Similar security analysis are also conducted in \cite{seds,seok2019,jiang2020}.}

\change{Thanks to the implementation of a symmetric encryption algorithm and a secure key agreement between the peers, \textit{confidentiality and integrity} of data transmitted in D2D communications are ensured. Thus, data remain confidential (i.e., not understandable by unauthorized parties) and integral (i.e., not modifiable by those who do not have the key).}

\change{As previously explained, a secret key is generated by both the D2D transmitter and receiver through the DHKE protocol and is used for the encryption/decryption of data sent over D2D links. The DHKE algorithm allows \textit{resistance to attacks by eavesdroppers}, since it ensures that the generated secret key is not intercepted by malicious users.}

\change{By employing the gNB as a trusted third party in the interactions between the D2D peers, the SeT-D2D protocol also ensures \textit{resistance to man-in-the-middle attacks}, which represent the DHKE's main vulnerability.}

\change{The \textit{authentication} of the nodes belonging to the multicast group is managed, in the SeT-D2D protocol, by using the procedures described by 3GPP in \cite{3gpp}. Furthermore, the application of the HMAC to the messages sent by the D2D peers to the gNB and vice versa guarantees the authentication and the integrity of the transmitted messages. HMAC envisages the implementation of a hash function to a particular combination of the message to be transmitted and the private key known only by the sender and recipient. In so doing, any modification to the transmitted message would be easily detectable by the receiver, for which the verification process of the message authentication would fail.}

\change{Moreover, the SeT-D2D protocol requires that the \textit{signature} is implemented by the message senders (i.e., gNB and relay nodes) in order to certify the origin of the data and guarantee the \textit{non-repudiation} of the transmitted messages.}

\change{Finally, the assurance of the \textit{privacy protection} of the users is obtained by transmitting their SUCI instead of the SUPI for their identification on the air interface.}

\section{Conclusions}
\label{sec:conclusions}

In this work, the Secure and Trust Device-to-Device (SeT-D2D) protocol has been defined in order to foster a trusted multicast service delivery in 5G-oriented networks through an improved version of the Conventional Multicast Scheme (CMS) aided by secure D2D communications. 
The selection of trustworthy and efficient D2D transmitters is the heart of our proposal, since both channel conditions of D2D links and trust parameters affect the procedure carried out to choose the best relay node for each D2D receiver. Furthermore, data sent via D2D communications are protected thanks to the implementation of an encryption algorithm, for which the keys are generated through a trusted use of the Diffie-Hellman Key Exchange (DHKE) protocol. 

In order to evaluate the performance of the proposed SeT-D2D protocol, a simulation campaign has been conducted by using the Matlab tool. 
The comparison with other protocols, which either do not use social trustworthiness to evaluate the trust of the nodes (i.e., Se-D2D) or do not take into consideration at all trustworthiness (i.e., D2D), shows that SeT-D2D allows a better management of network resources, ensuring more efficient data delivery. 
In fact, SeT-D2D effectively guarantees a proper selection of trustworthy D2D transmitters, thanks to the use of a trustworthiness model consistent with the actual nature of the network nodes.


\begin{thebibliography}  {99}

\bibitem{D2D} 
A. Asadi, Q. Wang, and V. Mancuso, ``A Survey on Device-to-Device Communications in Cellular Networks,'' IEEE Communications Surveys \& Tutorials, 16:4,  pp. 1801-19, April 2014.

\bibitem{cisco}
{CISCO. Cisco Visual Networking Index: Forecast and Methodology, 2016--2021}. 

\bibitem{embms} 
General aspects and principles for interfaces supporting Multimedia Broadcast Multicast Service (MBMS) within E-UTRAN, 3GPP TS 36.440 Release 11, 2012.

\bibitem{cms}
J. Liu, W. Chen, Z. Cao, and K. Letaief, ``Dynamic power and sub- carrier allocation for OFDMA-based wireless multicast systems," IEEE International Conference on Communications (ICC), May 2008. 

\bibitem{D2D-SF} 
L. Militano, M. Condoluci, G. Araniti, A. Molinaro, A. Iera, and G.-M. Muntean, ``Single Frequency-Based Device-to-Device-Enhanced Video Delivery for Evolved Multimedia Broadcast and Multicast Services," IEEE Transactions on Broadcasting, 61:2, pp. 263-278, March 2015.

\bibitem{d2d_multicast_1}
L. Feng, P. Zhao, F. Zhou, M. Yin, P. Yu, W. Li, and X. Qiu, ``Resource Allocation for 5G D2D Multicast Content Sharing in Social-Aware Cellular Networks," IEEE Communications Mag., 56:3, March 2018.

\bibitem{d2d_multicast_2}
H. Meshgi, D. Zhao, and R. Zheng, ``Optimal Resource Allocation in Multicast Device-to-Device Communications Underlaying LTE Networks," IEEE Transactions on Vehicular Technology, 66:9, Sept. 2017.

\bibitem{tvt}
F. Rinaldi, S. Pizzi, A. Orsino, A. Iera, A. Molinaro, and G. Araniti, ``A novel approach for MBSFN Area Formation aided by D2D Communications for eMBB Service Delivery in 5G NR Systems," IEEE Transactions on Vehicular Technology, Vol. 69, No. 2, Feb. 2020.

\bibitem{survey_d2d_sec1}
M. Wang, and Z. Yan, ``A Survey on Security in D2D Communications," Mobile Netw. and Appl. Journal", 22:2, pp.195-208, April 2017.

\bibitem{ss}
{ERICSSON White Paper, ``5G Security - Enabling a Trustworthy 5G System," 2018.} 

\bibitem{future_internet}
S. Pizzi, C. Suraci, L. Militano, A. Orsino, A. Molinaro, A. Iera, and G. Araniti, ``Enabling Trustworthy Multicast Wireless Services through D2D Communications in 5G Networks," Future Int. Journal, July 2018.

\bibitem{tob}
S. Pizzi, C. Suraci, A. Iera, A. Molinaro, and G. Araniti, ``A Sidelink-Aided Approach for Secure Multicast Service Delivery: from Human-Oriented Multimedia Traffic to Machine Type Communications," IEEE Transactions on Broadcasting, 2020.

\bibitem{3gpp}
Security Architecture and Procedures for 5G System, document 3GPP TS 33.501, Technical Specification Group Services and System Aspects.

\bibitem{5G_slicing}  
C. Campolo, A. Molinaro, A. Iera, and F. Menichella, ``5G Network Slicing for Vehicle-to-Everything Services," IEEE Wireless Communications, 24:6, pp. 38-45, Dec. 2017.

\bibitem{types_traffic}  
P. Popovski, K. F. Trillingsgaard, O. Simeone, and G. Durisi, ``5G Wireless Network Slicing for eMBB, URLLC, and mMTC: A Communication-Theoretic View," IEEE Access, April 2018.

\bibitem{5G-tec} 
I. F. Akyildiz et al., ``5G roadmap: 10 key enabling technologies," Elsevier Computer Networks, 106, pp. 17-48, 2016.

\bibitem{3gpp-arch}
System Architecture for the 5G System; Stage 2 (Release 15), document 3GPP TS 23.501, Technical Specification Group Services and System Aspects.

\bibitem{security-privacy} 
M. Haus, M. Waqas, A. Yi Ding, Yong Li, Sasu Tarkoma, and Jörg Ott, ``Security and Privacy in Device-to-Device (D2D) Communication: A Review," IEEE Comm. Surveys \& Tut., 19:2, pp. 1054-79, Jan. 2017. 

\bibitem{survey-security} 
M. Wang, and Z. Yan, ``A Survey on Security in D2D Communications," Mobile Networks and Applications, 22:2, pp. 195-208, 2017.

\bibitem{seds} 
A. Zhang, J. Chen, R. Qingyang Hu, and Yi Qian, ``SeDS: Secure Data Sharing Strategy for D2D Communication in LTE-Advanced Networks," IEEE Transactions on Vehicular Technology, 65:4, April 2016.

\bibitem{ometov}
S. Andreev, J. Hosek, T. Olsson, K. Johnsson, A. Pyattaev, A. Ometov, E. Olshannikova, M. Gerasimenko, P. Masek, Y. Koucheryavy, and T. Mikkonen, ``A unifying perspective on proximity-based cellular-assisted mobile social networking,'' IEEE Communications Magazine, 54:4, April 2016.

\bibitem{trust_indoor}  
M. Nitti, V. Popescu, and M. Fadda, ``Using an IoT Platform for Trustworthy D2D Communications in a Real Indoor Environment," IEEE Trans. on Netw. and Serv. Management, 16:1, pp. 234-245, Dec. 2018.

\bibitem{reliable_relay}  
F. H. Kumbhar, N. Saxena, and A. Roy, ``Reliable Relay: Autonomous Social D2D Paradigm for 5G LoS Communications,'' IEEE Communications Letters, 21:7, pp. 234-245, July 2017.

\bibitem{wang}
Bo Bai, Li Wang, Zhu Han, Wei Chen, and Tommy Svensson, ``Caching based socially-aware D2D communications in wireless content delivery networks: a hypergraph framework,''  IEEE Wireless Communications, 23:4, August 2016.

\bibitem{Can} 
A. B. Can, and B. Bhargava, ``SORT: A Self-ORganizing Trust Model for Peer-to-Peer Systems," IEEE Transactions on Dependable and Secure Computing, 10:1, Jan. 2013.

\bibitem{Chen} 
Z. Chen, R. Ling, C. Huang and Xu Zhu, ``A scheme of access service recommendation for the Social Internet of Things," International Journal of Communication Systems, 29:4, pp. 694-706, March 2016.

\bibitem{Peertrust} 
X. Li, and L. Liu, ``Peertrust: Supporting reputation-based trust for peer-to-peer electronic communities," IEEE Transactions on Knowledge and Data Engineering, 16:7, pp. 843-857, July 2004.

 \bibitem{Fuzzy} 
E. Damiani, S. De capitani di Vimercati, S. Paraboschi, M. Pesenti, P. Samarati, and S. Zara, ``Fuzzy logic techniques for reputation management in anonymous peer-to-peer systems," In EUSFLAT Conference, pp. 43-48, September 2003.

\bibitem{Scalable} 
F. Bao, R. Chen, and J. Guo, ``Scalable, adaptive and survivable trust management for community of interest based internet of things systems," in IEEE Eleventh International Symposium, 2013.

\bibitem{Adaptive} 
R. Chen, F. Bao, and J. Guo, ``Trust-based service management for social internet of things systems," IEEE Transactions on Dependable and Secure Computing, 13:6, pp. 684-696, Dec. 2016.

\bibitem{Subj} 
M. Nitti, R. Girau, L. Atzori, A. Iera, and G. Morabito, ``A Subjective Model for Trustworthiness Evaluation in the Social Internet of Things," in IEEE 23rd International Symposium on Personal Indoor and Mobile Radio Communications (PIMRC), 2012.

\bibitem{Obj} 
M. Nitti, R. Girau, and L. Atzori, ``Trustworthiness management in the social internet of things," IEEE Transactions on Knowledge and Data Engineering, 26:5, pp. 1253-1266, May 2014.

\bibitem{architecture_5G}  
G. Arfaoui, et al., ``A Security Architecture for 5G Networks," IEEE Access, 6, pp. 22466-22479, April 2018.   

\bibitem{jover19} 
R. P. Jover and V. Marojevic, ``Security and Protocol Exploit Analysis of the 5G Specifications," IEEE Access, v. 7, pp. 24956-63, Feb. 2019.

\bibitem{Jos1} 
A. J\o{}sang, R. Hayward, and S. Pope, ``Trust network analysis with subjective logic," In Proceedings of the 29th Australasian Computer Science Conference, pp. 85-94, 2006.

\bibitem{Iera1} 
L. Atzori, A. Iera, G. Morabito, and M. Nitti, ``The Social Internet of Things (SIoT) - When Social Networks meet the Internet of Things: Concept, Architecture and Network Characterization," Computer Networks, 56:16, pp. 3594-3608, Nov. 2012.

\bibitem{SWIM} 
S. Kosta, A. Mei and J. Stefa, ``Small World in Motion (SWIM): Modeling Communities in Ad-Hoc Mobile Networking," IEEE International Conference on Sensor, Mesh and Ad Hoc Communications and Networks (SECON 2010), Boston, MA, U.S.A., 2010.

\bibitem{onoff}
H. Alzaid, et al. ``Reputation-based trust systems for wireless sensor networks: A comprehensive review," IFIP International Conference on Trust Management. Springer, Berlin, Heidelberg, 2013.

\bibitem{seok2019} 
B. Seok, J. C. S. Sicato, T. Erzhena, C. Xuan, Y. Pan, and J. H. Park, ``Secure D2D Communication for 5G IoT Network Based on Lightweight Cryptography," Applied Sciences, 10:1, Dec. 2019.

\bibitem{jiang2020} 
Y. Jiang, Y. Shen, and Q. Zhu, ``A Lightweight Key Agreement Protocol Based on Chinese Remainder Theorem and ECDH for Smart Homes," Sensors, 20:5, March 2020.

\end{thebibliography}
\end{document}